\documentclass[12pt,aps,prd,showpacs,amsmath,amssymb]{revtex4}
\usepackage{graphicx}
\usepackage{epstopdf}
\usepackage{multirow}
\usepackage{subfigure}
\usepackage{appendix}
\input epsf
\begin{document}
\title{Partial decay widths of $P_{c}(4312)$ as a $\bar{D}\Sigma_{c}$ molecular state}
\author{Yong-Jiang Xu$^{1}$\footnote{xuyongjiang13@nudt.edu.cn}, Chun-Yu Cui$^2$, Yong-Lu Liu$^1$, and Ming-Qiu Huang$^{1,3}$\footnote{corresponding author: mqhuang@nudt.edu.cn}}
\affiliation{$^1$Department of Physics, College of Liberal Arts and Sciences, National University of Defense Technology , Changsha, 410073, Hunan, China}
\affiliation{$^2$Department of Physics, Third Military Medical University (Army Medical University), Chongqing, 400038, China}
\affiliation{$^3$Synergetic Innovation Center for Quantum Effects and Applications, Hunan Normal University, Changsha,  410081, Hunan, China}
\date{}
\begin{abstract}
In the present work, the partial decay widths of $P_{c}(4312)$ to $\eta_{c} p$ and $J/\psi p$ are investigated with the QCD sum rule method under the assumption that $P_{c}(4312)$ is a $\bar{D}\Sigma_{c}$ molecular state with $J^{P}=\frac{1}{2}^{-}$. In the analysis, the pole residue of $P_{c}(4312)$, one of the input parameters for the calculations of the strong decay constants, is calculated first. With the numerical values of the strong decay constants, the partial decay widths to $\eta_{c} p$ and $J/\psi p$ are estimated to be $\Gamma(P_{c}(4312)\rightarrow \eta_{c} p)=5.54^{+0.75}_{-0.5}\mbox{MeV}$ and $\Gamma(P_{c}(4312)\rightarrow J/\psi p)=1.67^{+0.92}_{-0.56}\mbox{MeV}$, respectively, which are compatible with the measured total width of $P_{c}(4312)$. The results suggest that it is reasonable to assign $P_{c}(4312)$ to be a $\bar{D}\Sigma_{c}$ molecular state with $J^{P}=\frac{1}{2}^{-}$.
\end{abstract}
\pacs{11.25.Hf,~ 11.55.Hx,~ 13.40.Gp.} \maketitle

\section{Introduction}\label{sec1}

Multiquark states with quark substructures $qq\bar{q}\bar{q}$, $qqqq\bar{q}$ and so on, are allowed both in the conventional quark model and quantum chromodynamics (QCD), the correct theory of the strong interaction. They provide a good platform for studying the nonperturbative behavior of QCD. Many physicists have focused on this topic since the observation of $X(3872)$ in 2003 by the Belle Collaboration \cite{belle1}, and there have been many theoretical and experimental progresses on the theme in the last decade (see review articles \cite{H.X.Chen} for details).

The pentaquark states, a typical kind of multiquark states, are the focus of research on the nonconventional hadrons, especially after the discoveries of the $P_{c}(4380)$ and $P_{c}(4450)$ states in 2015 by the LHCb Collaboration \cite{lhcb}. These studies based on different assumptions about the quark configurations of the hadrons, including meson-baryon molecules \cite{R.Chen,L.Roca,J.He,H.X.Huang,U.G.Meissner,C.W.Xiao,R.Chen1}, diquark-diquark-antiquark pentaquarks \cite{L.Maiani,V.V.Anisovich,R.Ghosh,Z.G.Wang}, compact diquark-triquark pentaquarks \cite{R.F.Lebed,R.L.Zhu}, the topological soliton model \cite{N.N.Scoccola}, genuine multiquark states other than molecules \cite{A.Mironov}, and kinematical effects related to the triangle singularity \cite{F.K.Guo,X.H.Liu,M.Mikhasenko}, etc.

Recently, a new pentaquark state $P_{c}(4312)$ with mass $m_{P_{c}(4312)}=4311.9\pm0.7^{+6.8}_{-0.6}\mbox{MeV}$ and total width $\Gamma_{P_{c}(4312)}=9.8\pm2.7^{+3.7}_{-4.5}\mbox{MeV}$ was discovered by the LHCb Collaboration in the $J/\psi p$ invariant mass spectrum of the $\Lambda_{b}\rightarrow J/\psi p K$ decay \cite{lhcb1}. Triggered by this observation, there are many theoretical investigations on the properties of this state through different approaches, such as QCD sum rule method \cite{M. Pavon,S.L.Zhu,J.R.Zhang,A. Pimikov}, potential models \cite{H.Huang,F.Giannuzzi,J.Ping,J.B.Cheng,J.He1} and so on \cite{P.Holma,C.W.Xiao1,B.Wang,M.Z.Liu,Z.H.Guo,C.J.Xiao}. However, the concrete nature and substructure of this state are not determined yet. More experimental and theoretical investigations are necessary to understand its properties. For example, studying its possible decay channels may provide valuable insights in this respect.

In this paper, we study the strong decay property of $P_{c}(4312)$ viewed as a $\bar{D}\Sigma_{c}$  molecular state with $J^{P}=\frac{1}{2}^{-}$ in the QCD sum rule method \cite{SVZ}. First, we calculate the pole residue of $P_{c}(4312)$, one of the input parameters when computing the strong decay constants. Then we turn to the strong decay constants of $P_{c}(4312)\rightarrow \eta_{c} p$ and $P_{c}(4312)\rightarrow J/\psi p$. With the above results, we give the partial decay widths, $\Gamma(P_{c}(4312)\rightarrow \eta_{c} p)=5.54^{+0.75}_{-0.5}\mbox{MeV}$ and $\Gamma(P_{c}(4312)\rightarrow J/\psi p)=1.67^{+0.92}_{-0.56}\mbox{MeV}$. The basic idea of the QCD sum rule method is that the correlation function of interpolating currents of hadrons can be represented in terms of hadronic parameters (the so-called hadronic side) and calculated at quark-gluon level by operator product expansion (OPE) (the so-called QCD side), and then by matching the two expressions we can extract the physical quantities of the considered hadron. The QCD sum rule method has extensively been used to investigate the X, Y, Z states which are candidates for the multiquark states; for a review, see Ref.\cite{R.M.Albuquerque}. It is reliable for us to investigate the ground pentaquark states using this method before more exact experiments are presented. In fact, there are some related works about the pentaquark states with the QCD sum rule method \cite{Z.G.Wang,H.X.Chen1,H.X.Chen2,H.X.Chen3,K.Azizi,Z.G.Wang1,J.R.Zhang}.

The rest of the paper is organized as follows. In Sec. \ref{sec2}, we give the sum rules for the pole residue of $P_{c}(4312)$ and the strong decay constants of $P_{c}(4312)\rightarrow \eta_{c} p$ and $P_{c}(4312)\rightarrow J/\psi p$. Section \ref{sec3} is devoted to the numerical analysis, and a short summary is given in Sec. \ref{sec4}. In Appendix \ref{appendix}, the spectral densities are shown.

\section{The derivation of the sum rules}\label{sec2}

In this section, the sum rules for the pole residue of $P_{c}(4312)$ and the strong decay constants of $P_{c}(4312)\rightarrow \eta_{c} p$ and $P_{c}(4312)\rightarrow J/\psi p$ are given.

\subsection{The pole residue}

To estimate the pole residue needed when calculating the strong decay constants, we start with the following two-point correlation function:
  \begin{equation}\label{2-point correlator}
 \Pi(p)=i\int d^{4}xe^{ipx}\langle0\mid\textsl{T}[J^{P_{c}}(x)\bar{J}^{P_{c}}(0)]\mid0\rangle
       =\not\!{p}\Pi_{1}(p^{2})+\Pi_{2}(p^{2}),
 \end{equation}
 where $J^{P_{c}}(x)$ is the interpolating current of $P_{c}(4312)$ considered as a $\bar{D}\Sigma_{c}$  molecular state with $J^{P}=\frac{1}{2}^{-}$ in the present work. According to Ref.\cite{H.X.Chen3}, $J^{P_{c}}(x)$ can take the form
 \begin{equation}\label{Pc interpotating current}
  J^{P_{c}}(x)=[\bar{c}(x)i\gamma_{5}d(x)][\epsilon^{abc}(u^{T}_{a}(x)C\gamma_{\mu}u_{b}(x))\gamma^{\mu}\gamma_{5}c_{c}(x)],
 \end{equation}
 where $T$ denotes the matrix transposition of the Dirac spinor indices, $C$ means charge conjugation matrix, and $a, b, c$ are color indices.

 There are three main steps in the QCD sum rule calculation which are as follows:
 \begin{itemize}
   \item{i} Presenting the correlation function in terms of hadronic parameters
   \item{ii} Calculating the correlator via OPE at the quark-gluon level
   \item{iii} Matching the two expressions with the help of quark-hadron duality and extracting the needed quantities
 \end{itemize}
In the last step, Borel transform is introduced to suppress the higher and continuum states' contributions and improve the convergence of the OPE series.

In order to express the two-point correlation function (\ref{2-point correlator}) physically, we insert a complete set of relevant states with the same quantum numbers as $J^{P_{c}}(x)$ between the two interpolating currents, isolate the ground-state term and finally get
 \begin{equation}\label{hadronic side 1}
 \Pi^{phe}(p)=\lambda^{2}_{P_{c}} \frac{\not\!{p}+m_{P_{c}}}{m^{2}_{P_{c}}-p^2}+\mbox{higher resonances},
 \end{equation}
 where $m_{P_{c}}$ is the hadronic mass, $\lambda_{P_{c}}$ is the pole residue of $P_{c}(4312)$ defined as $\langle0\mid J^{P_{c}}(0)\mid P_{c}(p,s)\rangle=\lambda_{P_{c}} u(p,s)$.

 On the other hand, $\Pi(p)$ can be calculated theoretically via OPE method at the quark-gluon level. To this end, one can insert the interpolating current $J^{P_{c}}(x)$ (\ref{Pc interpotating current}) into the correlation function (\ref{2-point correlator}), contract the relevant quark fields by Wick's theorem, and find
 \begin{eqnarray}
 \Pi^{OPE}(p)=&&-2i\epsilon_{abc}\epsilon_{a^{\prime}b^{\prime}c^{\prime}}\int d^{4}x e^{ipx}\gamma^{\mu}\gamma_{5}S^{(c)}_{cc^{\prime}}(x)\gamma^{\nu}\gamma_{5}\nonumber\\&&Tr[(i\gamma_{5})S^{(d)}_{dd^{\prime}}(x)(i\gamma_{5})S^{(c)}_{d^{\prime}d}(-x)]
 Tr[\gamma_{\mu}S^{(u)}_{bb^{\prime}}(x)\gamma_{\nu}CS^{(u)T}_{aa^{\prime}}(x)C],
 \end{eqnarray}
 where $S^{(c)}(x)$ and $S^{(q)}(x), q=u, d$ are the full charm- and up (down)-quark propagators, whose expressions are given in Appendix \ref{appendix1}. Through dispersion relation, $\Pi^{OPE}(p)$ can be written as
 \begin{equation}\label{OPE_1}
 \Pi^{OPE}(p)=\not\!{p}\int^{\infty}_{4m^{2}_{c}}ds\frac{\rho_{1}(s)}{s-p^2}
 +\int^{\infty}_{4m^{2}_{c}}ds\frac{\rho_{2}(s)}{s-p^2},
 \end{equation}
 where $\rho_{i}(s)=\frac{1}{\pi}\mbox{Im}\Pi^{OPE}_{i}(s), i=1,2$ are the spectral densities. The spectral density $\rho_{1}(s)$ is given in Appendix \ref{appendix}.

 Finally, we match the phenomenological side (\ref{hadronic side 1}) and the QCD representation (\ref{OPE_1}) for the Lorentz structure $\not\!{p}$,
 \begin{equation}
 \frac{\lambda^{2}_{P_{c}}}{m^{2}_{P_{c}}-p^2}+\mbox{higher resonances}=\int^{\infty}_{4m^{2}_{c}}ds\frac{\rho_{1}(s)}{s-p^2}.
 \end{equation}
 According to quark-hadron duality, the excited and continuum states' spectral density can be approximated by the QCD spectral density above some effective threshold $s^{P_{c}}_{0}$, whose value will be determined in Sec. \ref{sec3},
 \begin{equation}
 \frac{\lambda^{2}_{P_{c}}}{m^{2}_{P_{c}}-p^2}+\int^{\infty}_{s^{P_{c}}_{0}}ds\frac{\rho_{1}(s)}{s-p^2}+\mbox{subtractions}=\int^{\infty}_{4m^{2}_{c}}ds\frac{\rho_{1}(s)}{s-p^2}.
 \end{equation}
 Subtracting the contributions of the excited and continuum states, one gets
 \begin{equation}
 \frac{\lambda^{2}_{P_{c}}}{m^{2}_{P_{c}}-p^2}+\mbox{subtractions}=\int^{s^{P_{c}}_{0}}_{4m^{2}_{c}}ds\frac{\rho_{1}(s)}{s-p^2}.
 \end{equation}
 In order to eliminate the subtraction terms, it is necessary to make a Borel transform which can also improve the convergence of the OPE series and suppress the contributions from the excited and continuum states. As a result, we have
 \begin{equation}\label{2-point sum rule1}
 \lambda^{2}_{P_{c}}e^{-\frac{m^{2}_{P_{c}}}{M^{2}_{B}}}=\int^{s^{P_{c}}_{0}}_{4m^{2}_{c}}ds\rho_{1}(s)e^{-\frac{s}{M^{2}_{B}}},
 \end{equation}
 where $M^{2}_{B}$ is the Borel parameter. To get the sum rules for the mass and the pole residue $\lambda_{P_{c}}$, we take derivative of Eq.(\ref{2-point sum rule1}) with respect to $-\frac{1}{M^{2}_{B}}$ and divide it by the original expression. The final result is
 \begin{equation}
 m^{2}_{P_{c}}=(\frac{d}{d(-\frac{1}{M^{2}_{B}})}\int^{s^{P_{c}}_{0}}_{4m^{2}_{c}}ds\rho_{1}(s)e^{-\frac{s}{M^{2}_{B}}})/\int^{s_{0}}_{4m^{2}_{c}}ds\rho_{1}(s)e^{-\frac{s}{M^{2}_{B}}}.
 \end{equation}
 Substituting the obtained mass value into Eq.(\ref{2-point sum rule1}), we can give the sum rule of the pole residue $\lambda_{P_{c}}$. However, in the present case, the mass of $P_{c}(4312)$ is given by experiment. In order to improve the precision, we can substitute the experimental value of the mass in Eq.(\ref{2-point sum rule1}) to obtain the sum rule for the pole residue $\lambda_{P_{c}}$.

\subsection{The strong decay constants}

In the previous subsection, the sum rule of the pole residue of $P_{c}(4312)$ is given. We now turn to the calculation of the strong decay constants of $P_{c}(4312)\rightarrow \eta_{c} p$ and $P_{c}(4312)\rightarrow J/\psi p$. To this end, we begin with the following three-point correlation functions:
\begin{eqnarray}\label{3-point correlator}
&&\Gamma(p,p^{\prime},q)=i^{2}\int d^{4}xd^{4}y e^{ip^{\prime}x+iqy}\langle 0|T[J^{N}(x)J^{\eta_{c}}(y)\bar{J}^{P_{c}}(0)]|0\rangle,\nonumber\\
&&\Gamma_{\mu}(p,p^{\prime},q)=i^{2}\int d^{4}xd^{4}y e^{ip^{\prime}x+iqy}\langle 0|T[J^{N}(x)J^{J/\psi}_{\mu}(y)\bar{J}^{P_{c}}(0)]|0\rangle,
\end{eqnarray}
where $p=p^{\prime}+q$, $J^{P_{c}}(x)$ is the interpolating current of $P_{c}(4312)$ defined in (\ref{Pc interpotating current}), $J^{N}(x)$ , $J^{\eta_{c}}(x)$ and $J^{J/\psi}_{\mu}(x)$ are the interpolating currents of the proton, $\eta_{c}$ and $J/\psi$, respectively. The interpolating currents take the following form:
\begin{eqnarray}
&&J^{N}(x)=\epsilon_{abc}[u^{T}_{a}(x)C\gamma_{\mu}u_{b}(x)]\gamma_{5}\gamma^{\mu}d_{c}(x),\nonumber\\
&&J^{\eta_{c}}(x)=\bar{c}(x)i\gamma_{5}c(x),\nonumber\\
&&J^{J/\psi}_{\mu}(x)=\bar{c}(x)\gamma_{\mu}c(x),
\end{eqnarray}
where $T$ denotes the matrix transposition of the Dirac spinor indices, $C$ means charge conjugation, and $a, b, c$ are color indices.

Following the same procedures done above, we calculate the three-point correlators both phenomenologically and theoretically and extract the needed sum rules by matching the two representations of the correlation functions.

In order to get the physical representation of the three-point correlation functions (\ref{3-point correlator}), we insert complete sets of states having the same quantum numbers as the interpolating currents into the three-point correlation functions and define the following matrix elements:
\begin{eqnarray}\label{interpolating currents}
&&\langle 0|J^{N}|N(p^{\prime})\rangle=\lambda_{N}u^{N}(p^{\prime}),\nonumber\\
&&\langle 0|J^{J/\psi}_{\mu}|J/\psi(q)\rangle=f_{J/\psi}m_{J/\psi}\epsilon_{\mu}(q),\nonumber\\
&&\langle 0|J^{\eta_{c}}|\eta_{c}(q)\rangle=\frac{f_{\eta_{c}}m^{2}_{\eta_{c}}}{2m^{2}_{c}},
\end{eqnarray}
\begin{eqnarray}\label{matrix element}
&&\langle N(p^{\prime})\eta_{c}(q)|P_{c}(p)\rangle=ig\bar{u}^{N}(p^{\prime})u^{P_{c}}(p),\nonumber\\
&&\langle N(p^{\prime})J/\psi(q)|P_{c}(p)\rangle=\epsilon^{*}_{\mu}(q)\bar{u}^{N}(p^{\prime})(f_{1}\gamma^{\mu}-if_{2}\frac{\sigma^{\mu\nu}q_{\nu}}{m_{N}+m_{P_{c}}})\gamma_{5}u^{P_{c}}(p),
\end{eqnarray}
where $f_{\eta_{c}}$ and $m_{\eta_{c}}$ are the decay constant and mass of the $\eta_{c}$ state, $m_{J/\psi}$, $f_{J/\psi}$, and $\epsilon_{\mu}(q)$ are the mass, decay constant, and polarization vector of the $J/\psi$ state, $\lambda_{N}$ and $u^{N}(p^{\prime})$ are the residue and spinor of the proton, and $g$, $f_{1}$, and $f_{2}$ are the strong decay constants, respectively. After algebraic calculations, we reach the phenomenological side of the sum rules as follows:
\begin{equation}\label{3-point physical side 1}
\Gamma(p,p^{\prime},q)=[\frac{g\lambda_{N}\lambda_{P_{c}}f_{\eta_{c}}m^{2}_{\eta_{c}}(m_{N}+m_{P_{c}})}
{2m_{c}(m^{2}_{P_{c}}-p^{2})(m^{2}_{\eta_{c}}-q^{2})(m^{2}_{N}-p^{\prime2})}
+\frac{a}{(m^{2}_{\eta_{c}}-q^{2})(m^{2}_{N}-p^{\prime2})}]\not\!{p^{\prime}}+\cdots
\end{equation}
\begin{eqnarray}\label{3-point physical side}
\Gamma_{\mu}(p,p^{\prime},q)=&&[-\frac{\lambda_{N}\lambda_{P_{c}}f_{J/\psi}m_{J/\psi}}{(m^{2}_{P_{c}}-p^2)(m^{2}_{N}-p^{\prime 2})(m^{2}_{J/\psi}-q^2)}(f_{1}\frac{m_{N}+m_{P_{c}}}{m^{2}_{J/\psi}}
-f_{2}\frac{1}{m_{N}+m_{P_{c}}})\nonumber\\&&+\frac{a_{1}}{(m^{2}_{J/\psi}-q^{2})(m^{2}_{N}-p^{\prime2})}]\not\!{p^{\prime}}\not\!{q}\gamma_{5}q_{\mu}\nonumber\\&&+\{-\frac{\lambda_{N}\lambda_{P_{c}}f_{J/\psi}m_{J/\psi}}{(m^{2}_{P_{c}}-p^2)(m^{2}_{N}-p^{\prime 2})(m^{2}_{J/\psi}-q^2)}[-f_{1}(m_{N}+m_{P_{c}})+f_{2}\frac{q^{2}-p^{\prime2}}{m_{N}+m_{P_{c}}}]
\nonumber\\&&+\frac{a_{2}}{(m^{2}_{J/\psi}-q^{2})(m^{2}_{N}-p^{\prime2})}\}\not\!{p^{\prime}}\gamma_{\mu}\gamma_{5}+\cdots
\end{eqnarray}
where only the Lorentz structures $\not\!{p^{\prime}}$, $\not\!{p^{\prime}}\not\!{q}\gamma_{5}q_{\mu}$, and $\not\!{p^{\prime}}\gamma_{\mu}\gamma_{5}$ we are interested in, remained, and $a$, $a_{1}$, and $a_{2}$ are constant parameters introduced to parameterize the transitions between the ground states and the excited states similar to Ref.\cite{Z.G.Wang2}.

On the theoretical side, by inserting the interpolating currents (\ref{Pc interpotating current}) and (\ref{interpolating currents}) into the three-point correlation function (\ref{3-point correlator}) and contracting the quark fields, we obtain the following representation of the correlation functions:
\begin{eqnarray}\label{3-point OPE side}
\Gamma(p,p^{\prime},q)&&=i^{2}2\epsilon_{abc}\epsilon_{a^{\prime}b^{\prime}c^{\prime}}\int d^{4}xd^{4}y e^{ip^{\prime}x+iqy}\gamma_{5}\gamma^{\alpha}S^{(d)}_{cd^{\prime}}(x)i\gamma_{5}S^{(c)}_{d^{\prime}d}(-y)(i\gamma_{5})
S^{(c)}_{dc^{\prime}}(y)\gamma^{\beta}\gamma_{5}\nonumber\\&&Tr[\gamma_{\alpha}S^{(u)}_{bb^{\prime}}(x)
\gamma_{\beta}CS^{(u)T}_{aa^{\prime}}(x)C]\nonumber\\
&&=\Gamma(p^{2},p^{\prime2},q^{2})\not\!{p^{\prime}}+\cdots,
\end{eqnarray}
\begin{eqnarray}
\Gamma_{\mu}(p,p^{\prime},q)&&=i^{2}2\epsilon_{abc}\epsilon_{a^{\prime}b^{\prime}c^{\prime}}\int d^{4}xd^{4}y e^{ip^{\prime}x+iqy}\gamma_{5}\gamma^{\alpha}S^{(d)}_{cd^{\prime}}(x)i\gamma_{5}S^{(c)}_{d^{\prime}d}(-y)\gamma_{\mu}
S^{(c)}_{dc^{\prime}}(y)\gamma^{\beta}\gamma_{5}\nonumber\\&&Tr[\gamma_{\alpha}S^{(u)}_{bb^{\prime}}(x)
\gamma_{\beta}CS^{(u)T}_{aa^{\prime}}(x)C]\nonumber\\
&&=\Gamma_{1}(p^{2},p^{\prime2},q^{2})\not\!{p^{\prime}}\not\!{q}\gamma_{5}q_{\mu}
+\Gamma_{2}(p^{2},p^{\prime2},q^{2})\not\!{p^{\prime}}\gamma_{\mu}\gamma_{5}+\cdots,
\end{eqnarray}
where the coefficients $\Gamma(p^{2},p^{\prime2},q^{2})$, $\Gamma_{1}(p^{2},p^{\prime2},q^{2})$, and $\Gamma_{2}(p^{2},p^{\prime2},q^{2})$ can be represented as by the dispersion relation
\begin{equation}\label{3-point OPE side 1}
\Gamma_{i}(p^{2},p^{\prime2},q^{2})=\int^{\infty}_{4m^{2}_{c}}ds\int^{\infty}_{0}du\frac{\rho^{(3)}_{i}(p^{2},s,u)}
{(s-q^{2})(u-p^{\prime2})},
\end{equation}
where $\Gamma_{i}(p^{2},p^{\prime2},q^{2})$ stand for $\Gamma(p^{2},p^{\prime2},q^{2})$, $\Gamma_{1}(p^{2},p^{\prime2},q^{2})$, and $\Gamma_{2}(p^{2},p^{\prime2},q^{2})$, and $\rho^{(3)}_{i}(p^{2},s,u)$ are the corresponding spectral densities which are given in Appendix \ref{appendix}.

Matching the hadronic representations (\ref{3-point physical side 1}), (\ref{3-point physical side}) with the QCD representations (\ref{3-point OPE side 1}) for the corresponding Lorentz structures and using the quark-hadron duality, one has
\begin{eqnarray}
&&\frac{g\lambda_{N}\lambda_{P_{c}}f_{\eta_{c}}m^{2}_{\eta_{c}}(m_{N}+m_{P_{c}})}
{2m_{c}(m^{2}_{P_{c}}-p^{2})(m^{2}_{\eta_{c}}-q^{2})(m^{2}_{N}-p^{\prime2})}
+\frac{a}{(m^{2}_{\eta_{c}}-q^{2})(m^{2}_{N}-p^{\prime2})}\nonumber\\&&=
\int^{s_{\eta_{c}}}_{4m^{2}_{c}}ds\int^{u_{N}}_{0}du\frac{\rho^{(3)}(p^{2},s,u)}
{(s-q^{2})(u-p^{\prime2})},\nonumber\\
&&-\frac{\lambda_{N}\lambda_{P_{c}}f_{J/\psi}m_{J/\psi}}
{(m^{2}_{P_{c}}-p^{2})(m^{2}_{J/\psi}-q^{2})(m^{2}_{N}-p^{\prime2})}(f_{1}\frac{m_{N}+m_{P_{c}}}{m^{2}_{J/\psi}}
-f_{2}\frac{1}{m_{N}+m_{P_{c}}})\nonumber\\&&+\frac{a_{1}}{(m^{2}_{J/\psi}-q^{2})(m^{2}_{N}-p^{\prime2})}=\int^{s_{J/\psi}}_{4m^{2}_{c}}ds\int^{u_{N}}_{0}du\frac{\rho^{(3)}_{1}(p^{2},s,u)}
{(s-q^{2})(u-p^{\prime2})},\nonumber\\
&&-\frac{\lambda_{N}\lambda_{P_{c}}f_{J/\psi}m_{J/\psi}}
{(m^{2}_{P_{c}}-p^{2})(m^{2}_{J/\psi}-q^{2})(m^{2}_{N}-p^{\prime2})}[-f_{1}(m_{N}+m_{P_{c}})+f_{2}\frac{q^{2}-p^{\prime2}}{m_{N}+m_{P_{c}}}]
\nonumber\\&&+\frac{a_{2}}{(m^{2}_{J/\psi}-q^{2})(m^{2}_{N}-p^{\prime2})}=\int^{s_{J/\psi}}_{4m^{2}_{c}}ds\int^{u_{N}}_{0}du\frac{\rho^{(3)}_{2}(p^{2},s,u)}
{(s-q^{2})(u-p^{\prime2})},
\end{eqnarray}
where $s_{\eta_{c}}$, $s_{J/\psi}$, and $u_{N}$ are the threshold parameters corresponding to the $\eta_{c}$, $J/\psi$, and proton channels, respectively, whose values will be determined in Sec. \ref{sec3}.

Setting $p^{2}=q^{2}$ and doing double Borel transform: $p^{2}\rightarrow M^{2}_{B_{1}}$ and $p^{\prime2}\rightarrow M^{2}_{B_{2}}$, we get the following equations:
\begin{eqnarray}
&&g\lambda_{N}\lambda_{P_{c}}(m_{N}+m_{P_{c}})\frac{f_{\eta_{c}}m^{2}_{\eta_{c}}}{2m_{c}}
\frac{(e^{-m^{2}_{P_{c}}/M^{2}_{B_{1}}}-e^{-m^{2}_{\eta_{c}}/M^{2}_{B_{1}}})}{m^{2}_{\eta_{c}}-m^{2}_{P_{c}}}e^{-m^{2}_{N}/M^{2}_{B_{2}}}
\nonumber\\&&+ae^{-m^{2}_{\eta_{c}}/M^{2}_{B_{1}}}e^{-m^{2}_{N}/M^{2}_{B_{2}}}=
\int^{s_{\eta_{c}}}_{4m^{2}_{c}}ds\int^{u_{N}}_{0}du e^{-s/M^{2}_{B_{1}}}e^{u/M^{2}_{B_{2}}}\rho^{(3)}(s,u),
\nonumber\\
&&-\lambda_{N}\lambda_{P_{c}}f_{J/\psi}m_{J/\psi}[f_{1}\frac{m_{N}+m_{P_{c}}}{m^{2}_{J/\psi}}
\frac{e^{-m^{2}_{P_{c}}/M^{2}_{B_{1}}}-e^{-m^{2}_{J/\psi}/M^{2}_{B_{1}}}}{m^{2}_{J/\psi}-m^{2}_{P_{c}}}e^{-m^{2}_{N}/M^{2}_{B_{2}}}
\nonumber\\&&-f_{2}\frac{1}{m_{N}+m_{P_{c}}}\frac{e^{-m^{2}_{P_{c}}/M^{2}_{B_{1}}}-e^{-m^{2}_{J/\psi}/M^{2}_{B_{1}}}}{m^{2}_{J/\psi}-m^{2}_{P_{c}}}e^{-m^{2}_{N}/M^{2}_{B_{2}}}]
+a_{1}e^{-m^{2}_{J/\psi}/M^{2}_{B_{1}}}e^{-m^{2}_{N}/M^{2}_{B_{2}}}\nonumber\\&&=
\int^{s_{J/\psi}}_{4m^{2}_{c}}ds\int^{u_{N}}_{0}du e^{-s/M^{2}_{B_{1}}}e^{u/M^{2}_{B_{2}}}\rho^{(3)}_{1}(s,u),
\nonumber\\
&&-\lambda_{N}\lambda_{P_{c}}f_{J/\psi}m_{J/\psi}[-f_{1}(m_{N}+m_{P_{c}})
\frac{(e^{-m^{2}_{P_{c}}/M^{2}_{B_{1}}}-e^{-m^{2}_{J/\psi}/M^{2}_{B_{1}}})}{m^{2}_{J/\psi}-m^{2}_{P_{c}}}e^{-m^{2}_{N}/M^{2}_{B_{2}}}
\nonumber\\&&+\frac{f_{2}}{m_{N}+m_{P_{c}}}(-e^{-m^{2}_{P_{c}}/M^{2}_{B_{1}}}e^{-m^{2}_{N}/M^{2}_{B_{2}}}
+\frac{m^{2}_{N}+m^{2}_{J/\psi}}{m^{2}_{J/\psi}-m^{2}_{P_{c}}}(e^{-m^{2}_{P_{c}}/M^{2}_{B_{1}}}-e^{-m^{2}_{J/\psi}/M^{2}_{B_{1}}})e^{-m^{2}_{N}/M^{2}_{B_{2}}})]
\nonumber\\&&+a_{2}e^{-m^{2}_{J/\psi}/M^{2}_{B_{1}}}e^{-m^{2}_{N}/M^{2}_{B_{2}}}\nonumber\\&&=
\int^{s_{J/\psi}}_{4m^{2}_{c}}ds\int^{u_{N}}_{0}du e^{-s/M^{2}_{B_{1}}}e^{u/M^{2}_{B_{2}}}\rho^{(3)}_{2}(s,u).
\end{eqnarray}
Taking derivative of the above equations with respect to $-1/M^{2}_{B_{1}}$ and solving related equations, we obtain the sum rules of the strong decay constants as follows:
\begin{equation}
g=\frac{2m_{c}}{\lambda_{N}\lambda_{P_{c}}f_{\eta_{c}}m^{2}_{\eta_{c}}(m_{N}+m_{P_{c}})}
e^{m^{2}_{P_{c}}/M^{2}_{B_{1}}}e^{m^{2}_{N}/M^{2}_{B_{2}}}A(M^{2}_{B_{1}},M^{2}_{B_{2}},s_{\eta_{c}},u_{N}),
\end{equation}
\begin{eqnarray}
f_{1}=&&\frac{m_{J/\psi}}{\lambda_{N}\lambda_{P_{c}}f_{J/\psi}(m_{P_{c}}+m_{N})(m^{2}_{J/\psi}+m^{2}_{N}-m^{2}_{P_{c}})}
e^{m^{2}_{P_{c}}/M^{2}_{B_{1}}}e^{m^{2}_{N}/M^{2}_{B_{2}}}\nonumber\\&&[(m^{2}_{P_{c}}-m^{2}_{N})A_{1}(M^{2}_{B_{1}},M^{2}_{B_{2}},s_{J\psi},u_{N})
+A_{2}(M^{2}_{B_{1}},M^{2}_{B_{2}},s_{J\psi},u_{N})],
\end{eqnarray}
\begin{eqnarray}
f_{2}=&&\frac{m_{N}+m_{P_{c}}}{\lambda_{N}\lambda_{P_{c}}f_{J/\psi}m_{J/\psi}(m^{2}_{J/\psi}+m^{2}_{N}-m^{2}_{P_{c}})}
e^{m^{2}_{P_{c}}/M^{2}_{B_{1}}}e^{m^{2}_{N}/M^{2}_{B_{2}}}\nonumber\\&&[m^{2}_{J/\psi}A_{1}(M^{2}_{B_{1}},M^{2}_{B_{2}},s_{J\psi},u_{N})
+A_{2}(M^{2}_{B_{1}},M^{2}_{B_{2}},s_{J\psi},u_{N})],
\end{eqnarray}
with
\begin{eqnarray}
&&A(M^{2}_{B_{1}},M^{2}_{B_{2}},s_{\eta_{c}},u_{N})=\int^{s_{\eta_{c}}}_{4m^{2}_{c}}ds\int^{u_{N}}_{0}du e^{-s/M^{2}_{B_{1}}}e^{u/M^{2}_{B_{2}}}[m^{2}_{\eta_{c}}\rho^{(3)}(s,u)-s\rho^{(3)}(s,u)-\frac{\partial\rho^{(3)}(s,u)}{\partial(-1/M^{2}_{B_{1}})}],
\nonumber\\&&
A_{1}(M^{2}_{B_{1}},M^{2}_{B_{2}},s_{J/\psi},u_{N})=\int^{s_{J/\psi}}_{4m^{2}_{c}}ds\int^{u_{N}}_{0}du e^{-s/M^{2}_{B_{1}}}e^{u/M^{2}_{B_{2}}}[m^{2}_{J/\psi}\rho^{(3)}_{1}(s,u)-s\rho^{(3)}_{1}(s,u)-\frac{\partial\rho^{(3)}_{1}(s,u)}{\partial(-1/M^{2}_{B_{1}})}],
\nonumber\\&&
A_{2}(M^{2}_{B_{1}},M^{2}_{B_{2}},s_{J/\psi},u_{N})=\int^{s_{J/\psi}}_{4m^{2}_{c}}ds\int^{u_{N}}_{0}du e^{-s/M^{2}_{B_{1}}}e^{u/M^{2}_{B_{2}}}[m^{2}_{J/\psi}\rho^{(3)}_{2}(s,u)-s\rho^{(3)}_{2}(s,u)-\frac{\partial\rho^{(3)}_{2}(s,u)}{\partial(-1/M^{2}_{B_{1}})}].
\end{eqnarray}

\section{Numerical analysis and The partial decay widths}\label{sec3}

The QCD sum rules for the pole residue and the strong decay constants contain some fundamental inputs which are presented in Table \ref{input parameters}. Besides these parameters, there are a few auxiliary parameters introduced during the calculations: the continuum thresholds and the Borel parameters. They are not physical quantities; hence, the physical observables should be approximately insensitive to them. Therefore, we look for working regions of these parameters such that the dependence of the physical quantities on these parameters is weak. The continuum thresholds are related to the square of the first exited states having the same quantum numbers as the interpolating currents, while the Borel parameters are determined by demanding that both the contributions of the higher states and continuum are sufficiently suppressed and the contributions coming from higher dimensional operators are small.
\begin{table}[htb]
\caption{Some input parameters needed in the calculations.}\label{input parameters}
\begin{tabular}{|c|c|}
  \hline
  Parameter      &   Value    \\
  \hline
  {$\langle\bar{q}q\rangle$}  &      $-(0.24\pm0.01)^{3}\mbox{GeV}^{3}$                     \\
  {$\langle g_{s}\bar{q}\sigma Gq\rangle$} & $(0.8\pm0.1)\langle\bar{q}q\rangle\mbox{GeV}^{2}$ \\
  {$\langle g^{2}_{s}GG\rangle$}    &     $0.88\pm0.25\mbox{GeV}^{4}$                \\
  {$m_{c}$}  &    $1.275^{+0.025}_{-0.035}\mbox{GeV}$\cite{M.Tanabashi}                   \\
  {$m_{J/\psi}$} & $3096.900\pm0.006\mbox{MeV}$\cite{M.Tanabashi}                \\
  {$m_{N}$}   &    $938.272081\pm0.000006\mbox{MeV}$\cite{M.Tanabashi}  \\
  {$m_{\eta_{c}}$} & $2.9839\pm0.5\mbox{GeV}$\cite{M.Tanabashi}\\
  {$\lambda^{2}_{N}$} & $0.0011\pm0.0005\mbox{GeV}^{6}$\cite{K.Azizi1}      \\
  {$f_{J/\psi}$} &  $481\pm36\mbox{MeV}$\cite{E.V.Veliev}    \\
  {$f_{\eta_{c}}$} & $0.387\pm0.007\mbox{GeV}$\cite{D.Becirevic}       \\
  \hline
\end{tabular}
\end{table}

We define two quantities, the ratio of the pole contribution to the total contribution (RP) and the ratio of the highest-dimensional term in the OPE series to the total OPE series (RH), as follows:
\begin{eqnarray}
&&RP\equiv\frac{\int^{s^{P_{c}}_{0}}_{4m^{2}_{c}}ds\rho_{1}(s)e^{-\frac{s}{M^{2}_{B}}}}{\int^{\infty}_{4m^{2}_{c}}ds\rho_{1}(s)e^{-\frac{s}{M^{2}_{B}}}},
\nonumber\\&&RH\equiv\frac{\int^{s^{P_{c}}_{0}}_{4m^{2}_{c}}ds\rho^{\langle\bar{q}q\rangle^{3}}_{1}(s)e^{-\frac{s}{M^{2}_{B}}}}{\int^{s^{P_{c}}_{0}}_{4m^{2}_{c}}ds\rho_{1}(s)e^{-\frac{s}{M^{2}_{B}}}}
\end{eqnarray}
for the two-point correlation function, and similar quantities for the three-point correlation functions.

\begin{figure}[htb]
\subfigure[]{
\includegraphics[width=7cm]{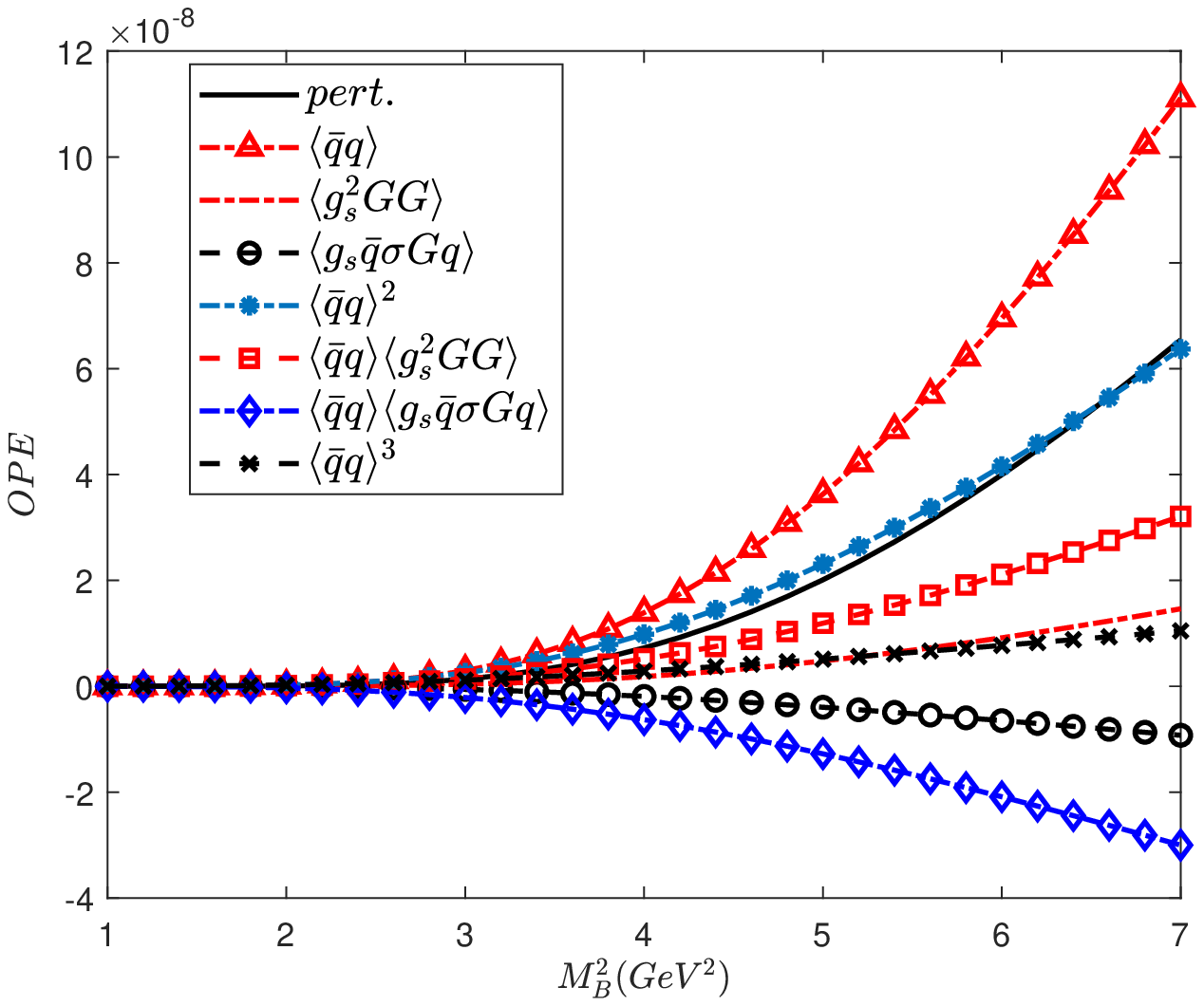}}
\subfigure[]{
\includegraphics[width=7cm]{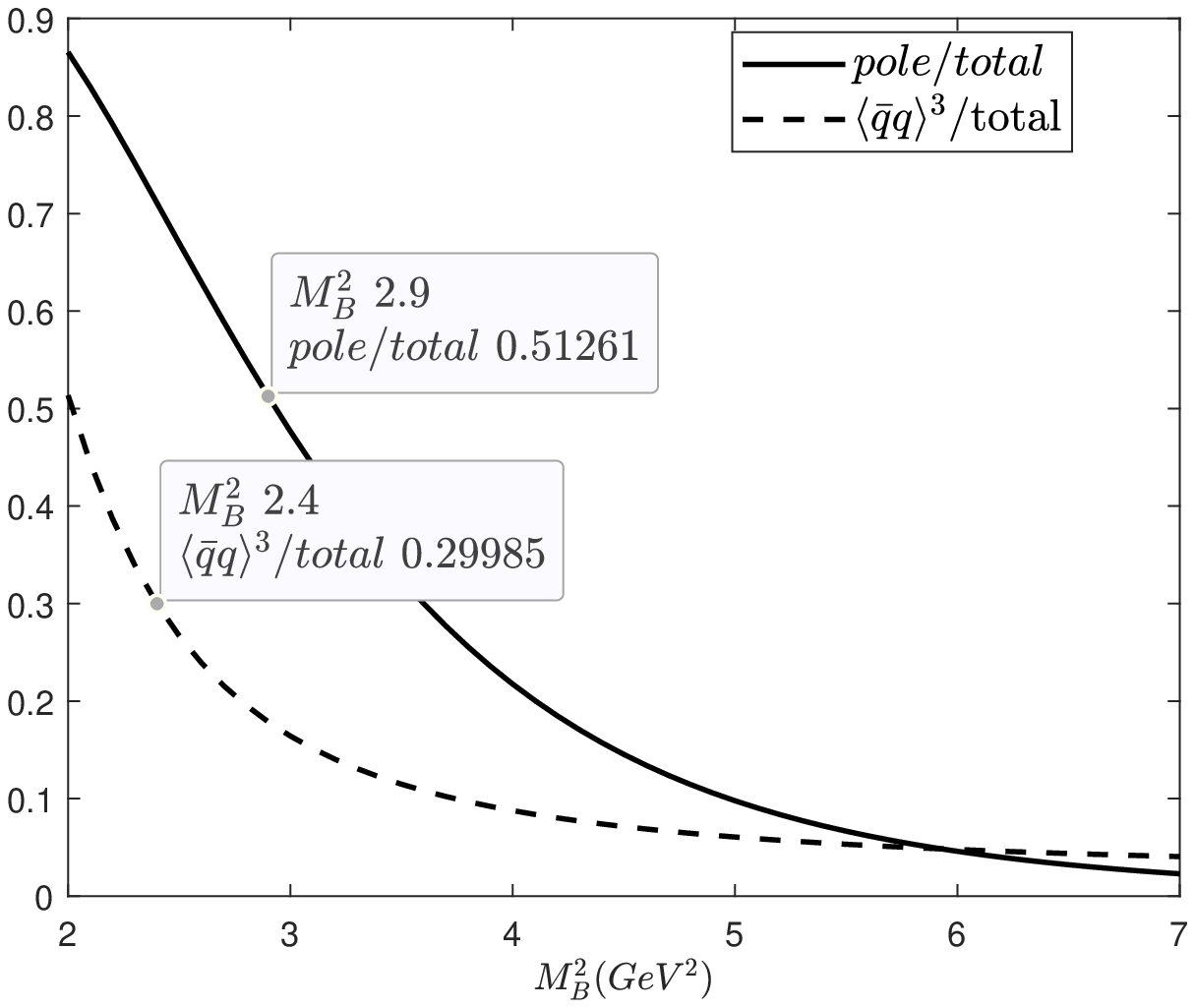}}
\caption{(a) denotes the various OPE contributions as functions of $M^{2}_{B}$ with $\sqrt{s^{P_{c}}_{0}}=4.8\mbox{GeV}$ and (b) represents $RP$ and $RH$ varying with $M^{2}_{B}$ at $\sqrt{s^{P_{c}}_{0}}=4.8\mbox{GeV}$.}\label{MB_range}
\end{figure}

We first analyze the pole residue $\lambda_{P_{c}}$. In Fig.\ref{MB_range}(a), we compare the various OPE contributions as functions of $M^{2}_{B}$ with $\sqrt{s^{P_{c}}_{0}}=4.8\mbox{GeV}$. From it, one can see that the quark condensate $\langle\bar{q}q\rangle$, the quark-gluon mixed condensate $\langle g_{s}\bar{q}\sigma Gq\rangle$, the four-quark condensate $\langle\bar{q}q\rangle^{2}$, and the dimension-8 term $\langle\bar{q}q\rangle g_{s}\bar{q}\sigma Gq\rangle$ play an important role in the OPE series, but they have opposite sign and cancel each other. As a result, the perturbative part still dominates the OPE series. Indeed, the highest-dimensional term $\langle\bar{q}q\rangle^{3}$ in our OPE is small relative to others. In other words, the OPE series is under control. Figure \ref{MB_range}(b) shows $RP$ and $RH$ varying with $M^{2}_{B}$ at $\sqrt{s^{P_{c}}_{0}}=4.8\mbox{GeV}$. The figure shows that it is needed to limit $M^{2}_{B}$ from $2.4\mbox{GeV}^{2}$ to $2.9\mbox{GeV}^{2}$ in order to simultaneously satisfy the requirements of pole dominance at the phenomenological side(the pole contribution is bigger than the continuum contribution) and convergence of the operator product expansion(the contribution from the highest-dimensional term is about $30$ percent of the total OPE series).

\begin{figure}[htb]
\includegraphics[width=10cm]{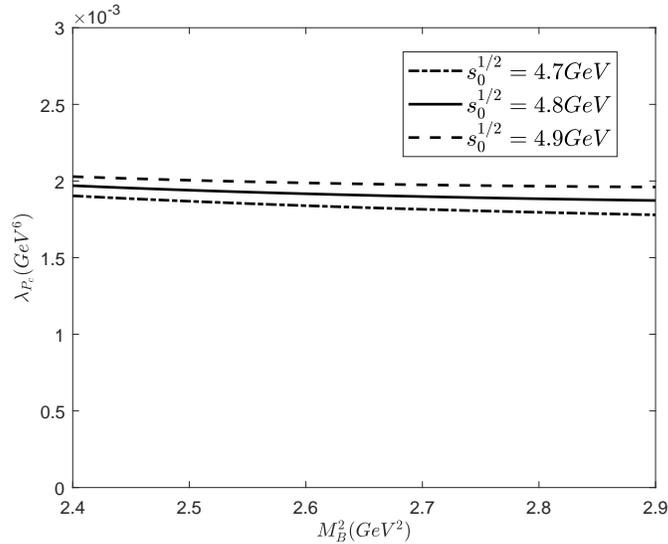}
\caption{The figure shows the dependence of the pole residue $\lambda_{P_{c}}$ on the Borel parameter $M^{2}_{B}$ in the determined interval at three different values of $s^{P_{c}}_{0}$.}\label{lambda}
\end{figure}

 With the obtained interval of $M^{2}_{B}$ and the experimental value of the mass $m_{P_{c}(4312)}=4311.9\pm0.7^{+6.8}_{-0.6}\mbox{MeV}$, the pole residue can be estimated. The result is represented in Fig.\ref{lambda}, from which it is obvious that the pole residue varies weakly with the parameters $s^{P_{c}}_{0}$ and $M^{2}_{B}$ in the interval determined above. As a result, we can reliably read the value of the pole residue, $\lambda_{P_{c}}=1.91^{+0.12}_{-0.13}\times10^{-3}\mbox{GeV}^{6}$.

\begin{figure}[htb]
\subfigure[]{
\includegraphics[width=7cm]{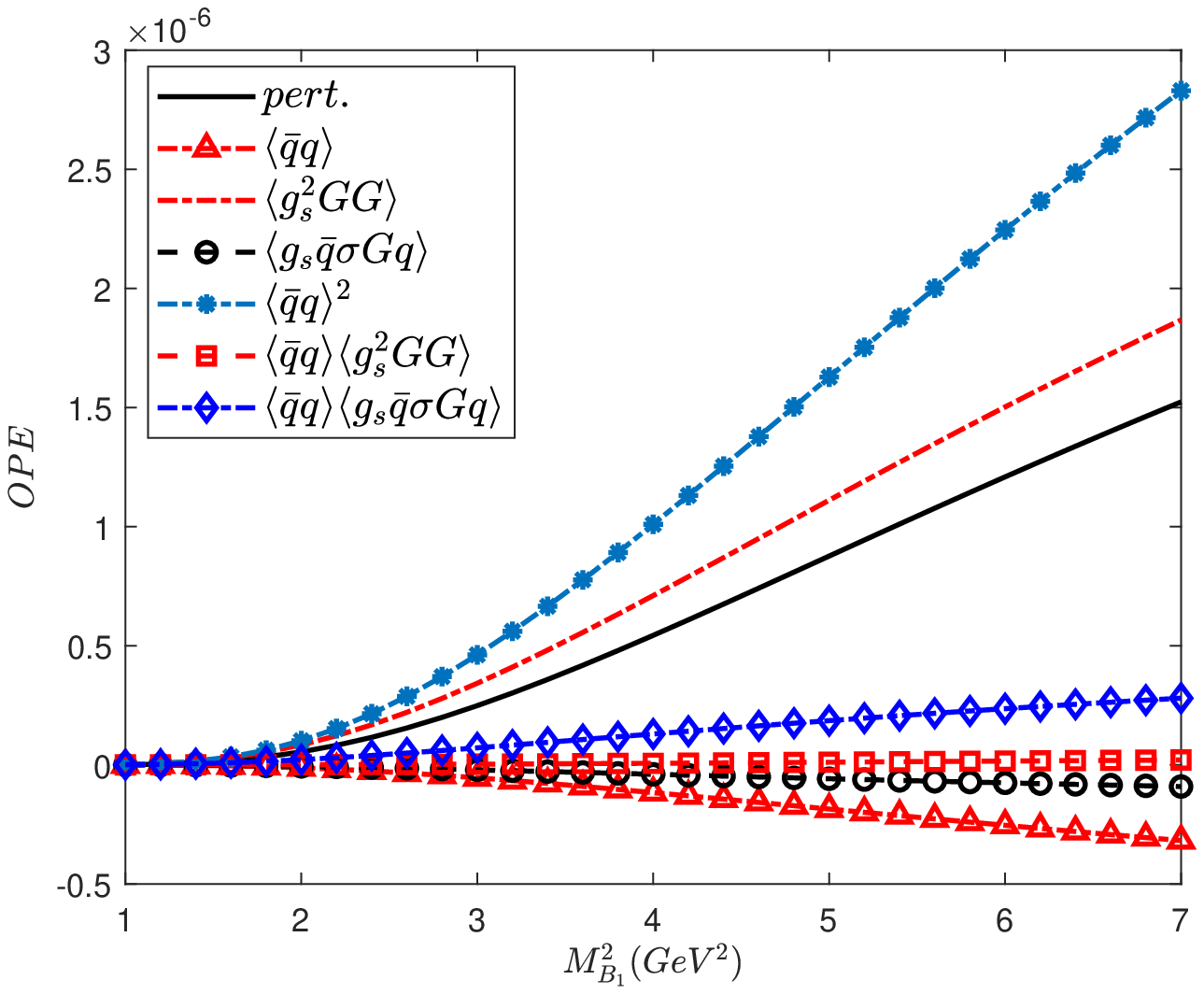}}
\subfigure[]{
\includegraphics[width=7cm]{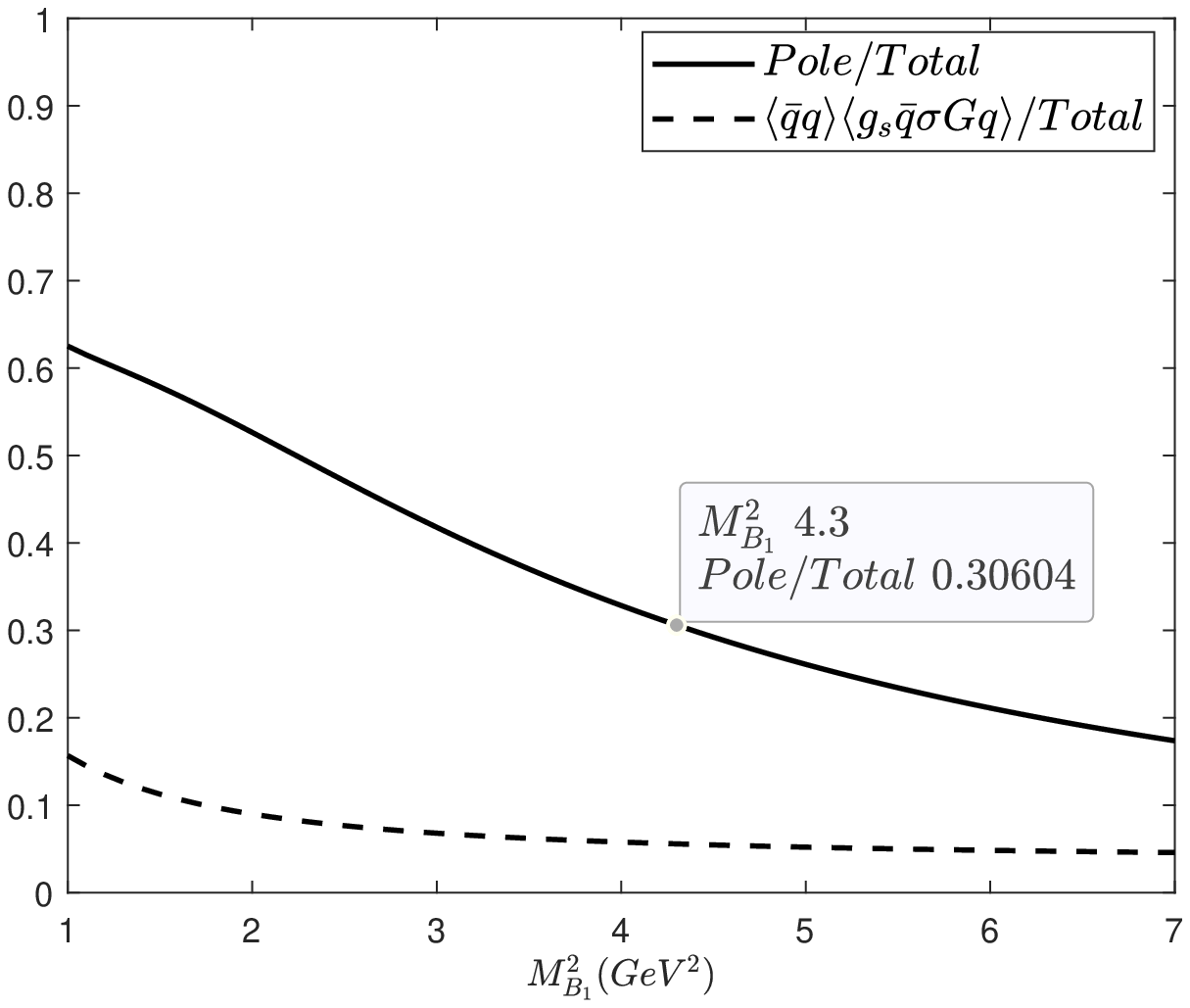}}
\subfigure[]{
\includegraphics[width=7cm]{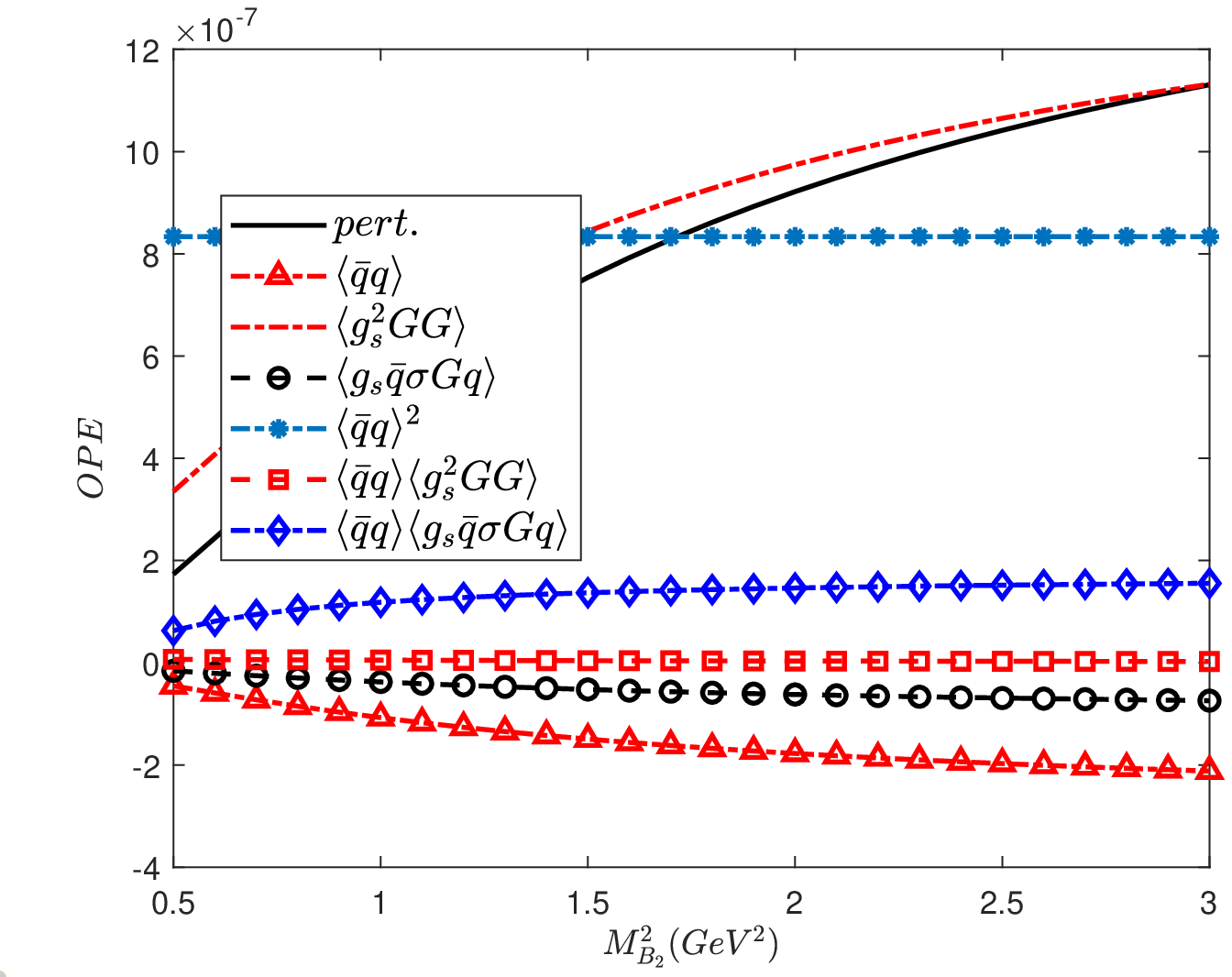}}
\subfigure[]{
\includegraphics[width=7cm]{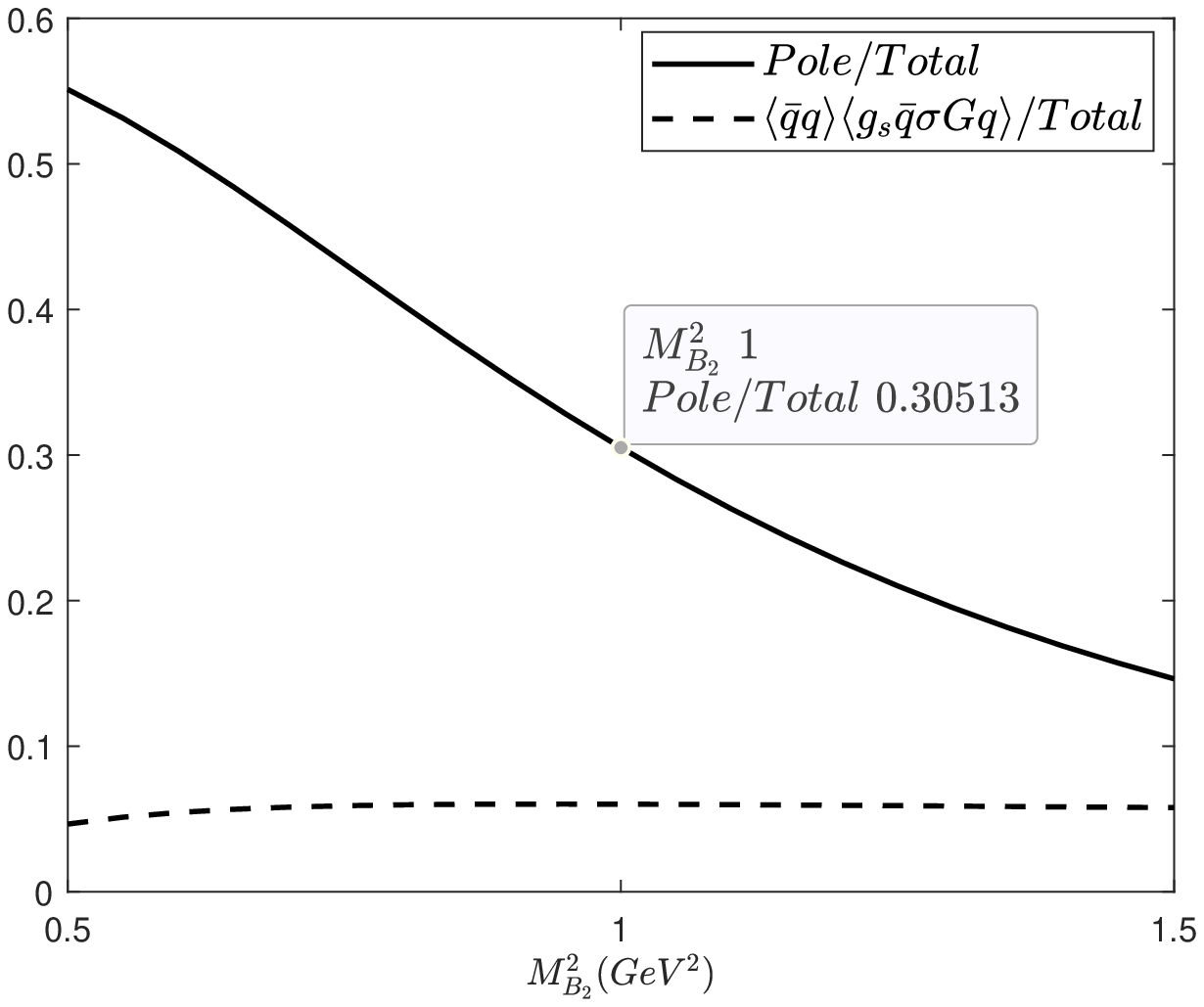}}
\caption{The coefficients of the Lorentz structure $\not\!{p^{\prime}}$ of the correlation function $\Gamma(p,p^{\prime},q)$, RP and RH, as functions of the Borel parameters $M^{2}_{B_{1}}$ with $M^{2}_{B_{2}}=0.9\mbox{GeV}^{2}$ are showed in (a) and (b) respectively. (c) and (d) represent the same quantities as functions of the Borel parameters $M^{2}_{B_{2}}$ with $M^{2}_{B_{1}}=3.7\mbox{GeV}^{2}$.}\label{etac-3-point OPE}
\end{figure}

\begin{figure}[htb]
\subfigure[]{
\includegraphics[width=7cm]{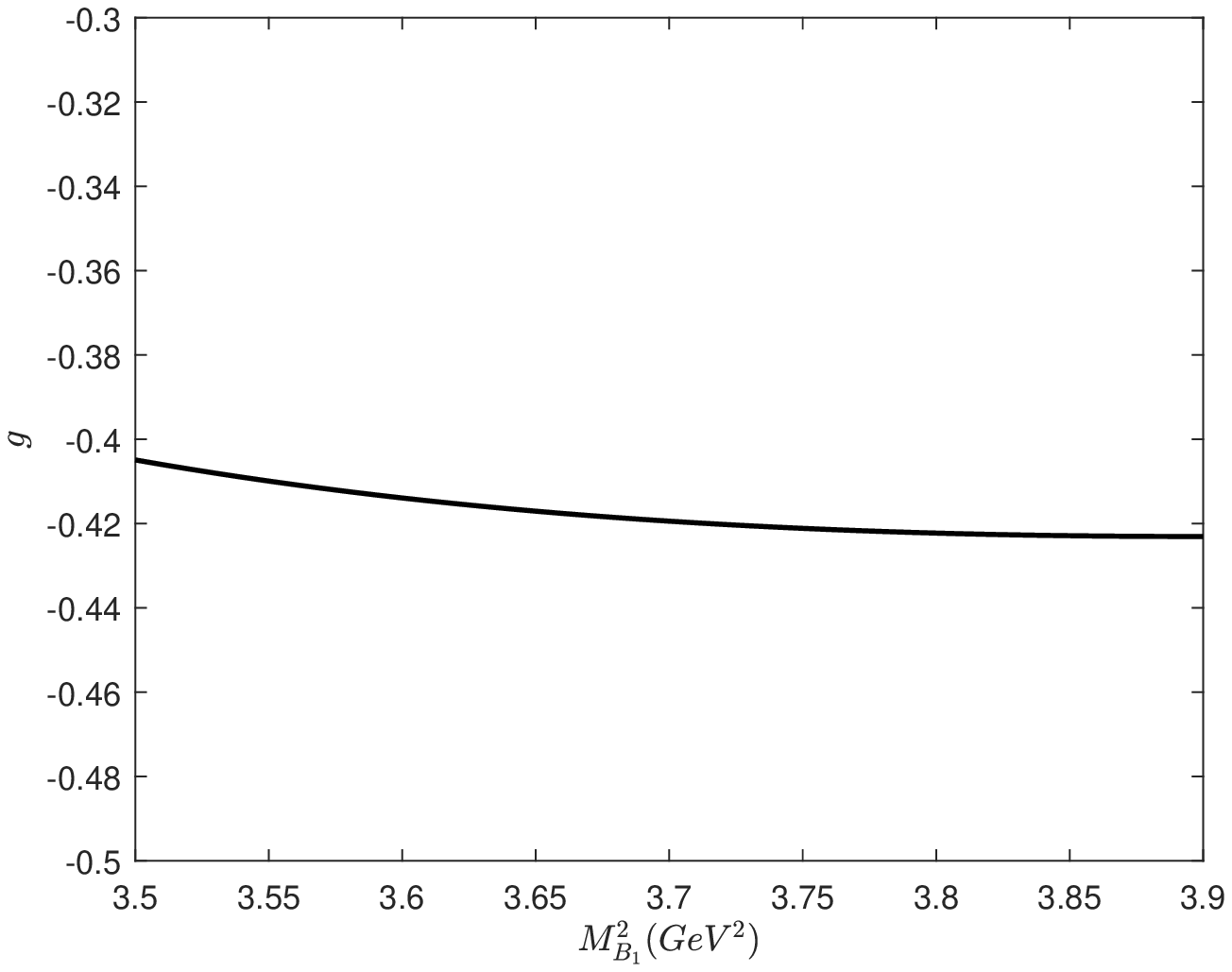}}
\subfigure[]{
\includegraphics[width=7cm]{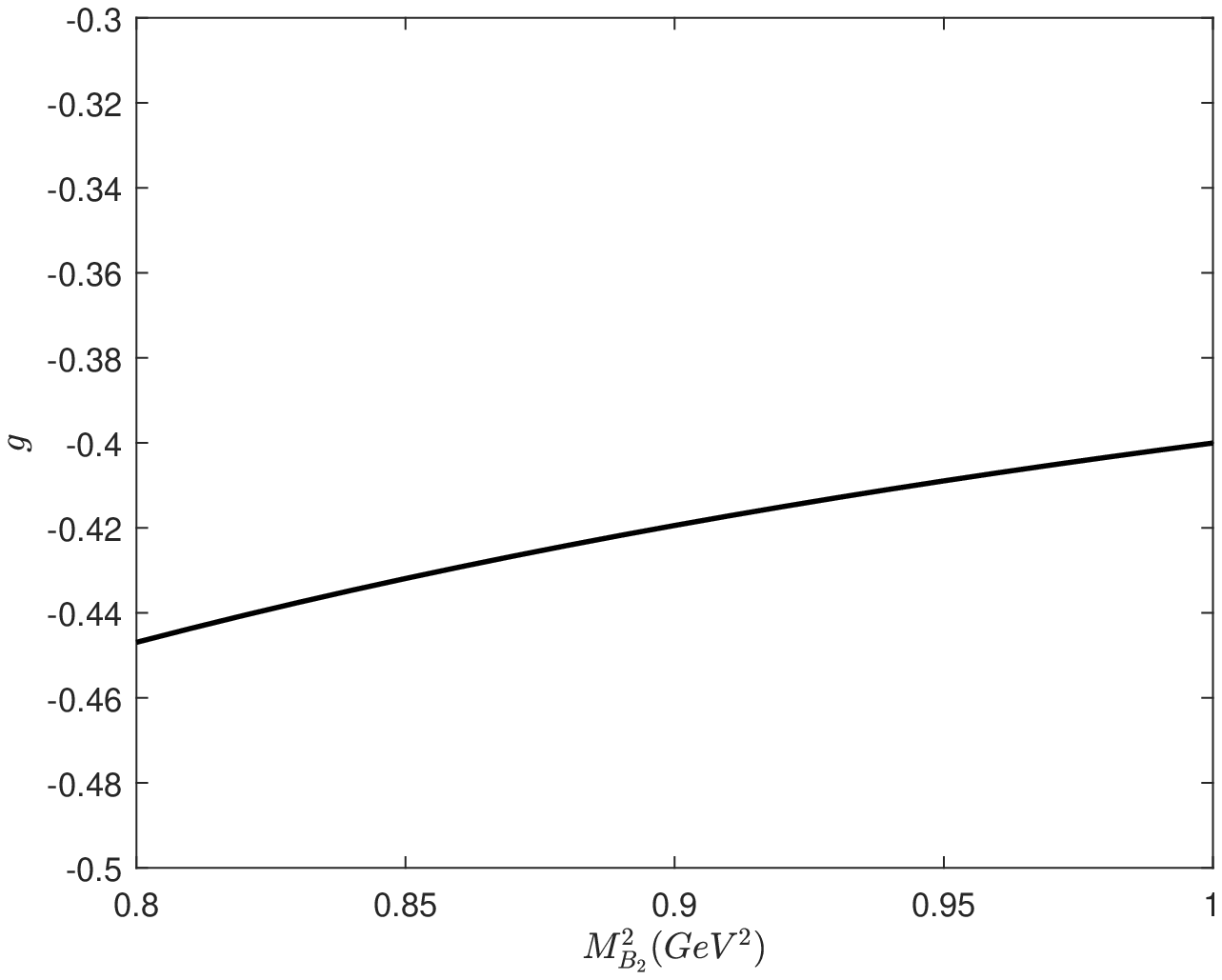}}
\caption{(a) and (b) show the strong decay constants of the decay $P_{c}(4312)\rightarrow\eta_{c} p$ in the allowed intervals of the Borel parameters $M^{2}_{B_{1}}$ and $M^{2}_{B_{2}}$, respectively.}\label{etac-coupling}
\end{figure}

Now, it is time to study the strong decay constants $g$ of the strong decay $P_{c}(4312)\rightarrow\eta_{c} p$, $f_{1}$ and $f_{2}$ of the strong decay $P_{c}(4312)\rightarrow J/\psi p$. Similar to above, we determine first the allowed ranges of the Borel parameters $M^{2}_{B_{1}}$ and $M^{2}_{B_{2}}$. To this end, we show the various OPE contributions of the Lorentz structure $\!\not\!p^{\prime}$ of the correlation function $\Gamma(p,p^{\prime},q)$ in Fig.\ref{etac-3-point OPE}(a), $RP$ and $RH$ in Fig.\ref{etac-3-point OPE}(b) as functions of $M^{2}_{B_{1}}$ with $M^{2}_{B_{2}}=0.9\mbox{GeV}^{2}$. Figures \ref{etac-3-point OPE}(c) and \ref{etac-3-point OPE}(d) depict the same quantities as functions of $M^{2}_{B_{2}}$ at $M^{2}_{B_{1}}=3.7\mbox{GeV}^{2}$. In the case of three-point correlation functions, as stated in Ref.\cite{M. Eidemuller}, the contributions of the pole-continuum transition terms may be larger than or the same order as the pole contribution and should not be neglected. In the present case, if we require the contribution from the pole larger than the continuum contribution, it is impossible to obtain suitable intervals of the Borel parameters. Therefore, we require that the pole term accounts for $30\%$ of the total contribution. Besides the above requirement, the Borel parameters are also constrained by the criterion that the physical quantities should be independent on the Borel parameters. Finally, the results are shown in Fig.\ref{etac-coupling}, from which we can see that the strong decay constant $g$ varies weakly with the Borel parameters and we can read the value of $g$: $g=-0.419^{+0.019}_{-0.028}$. In the above analysis, we take $s_{\eta_{c}}=3.5^{2}\mbox{GeV}^{2}$ and $u_{N}=1.7\mbox{GeV}^{2}$.

\begin{figure}[htb]
\subfigure[]{
\includegraphics[width=7cm]{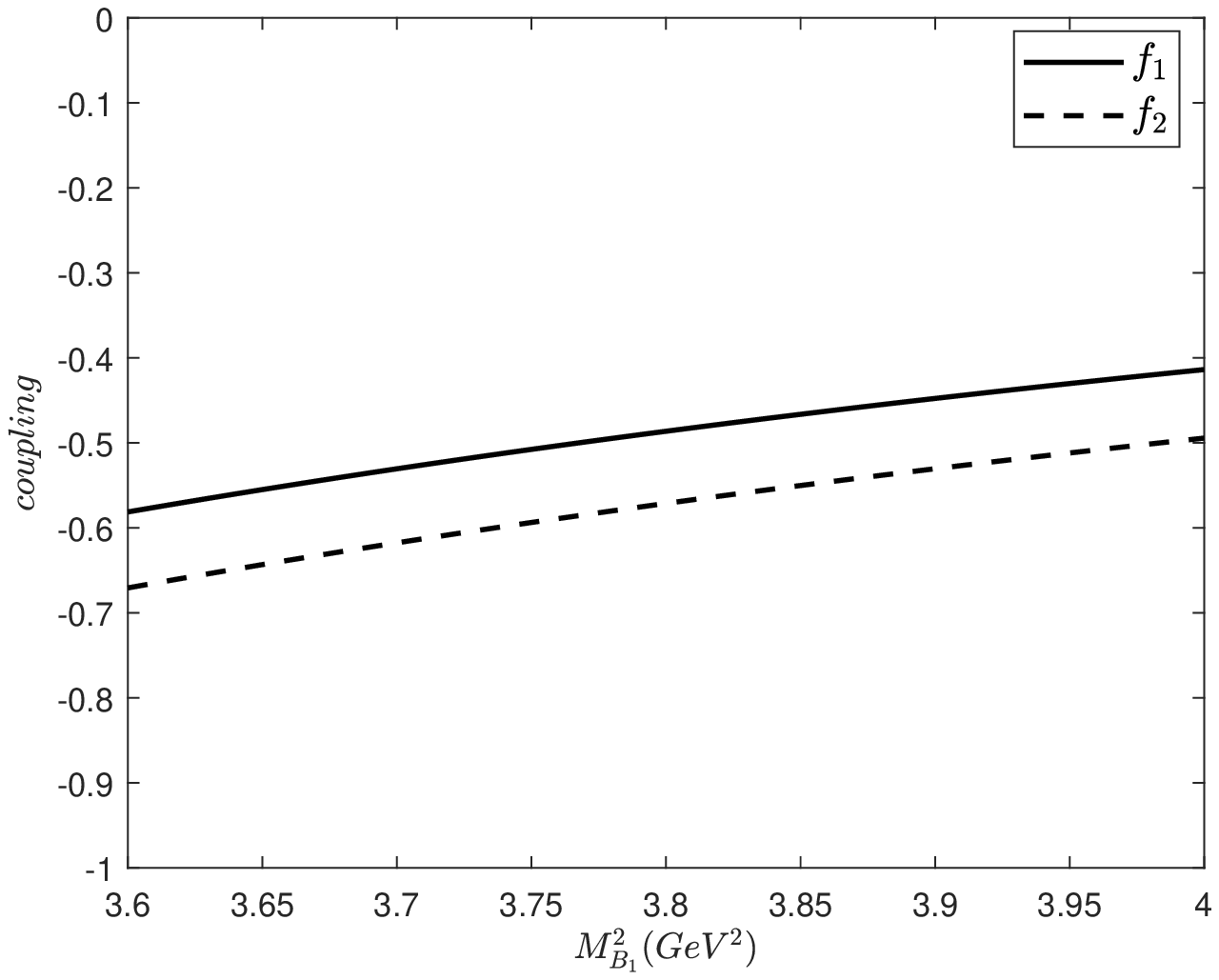}}
\subfigure[]{
\includegraphics[width=7cm]{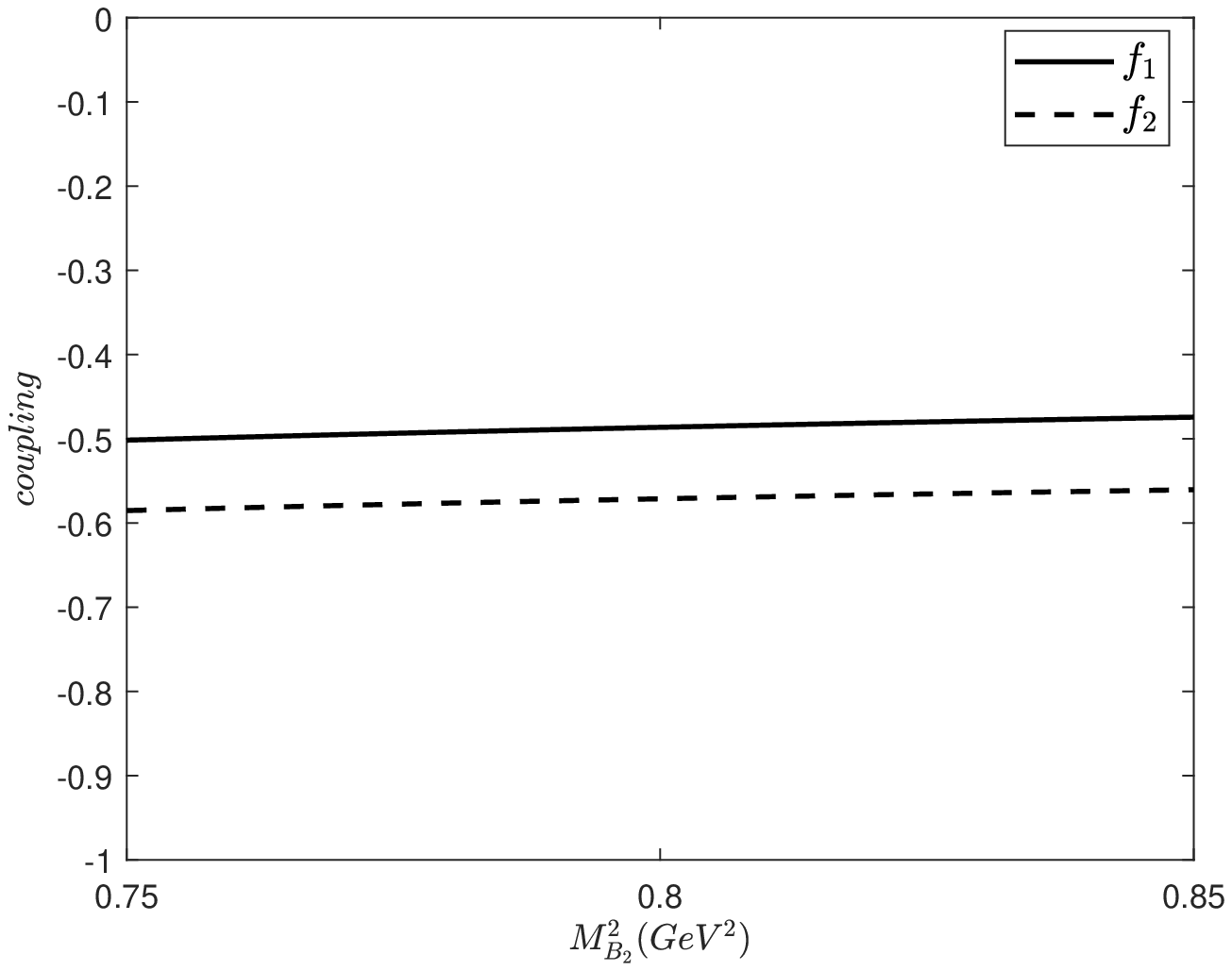}}
\caption{(a) and (b) shows the dependence of the sum rules for the strong decay constants $f_{1}$ and $f_{2}$ on the Borel parameters $M^{2}_{B_{1}}$ and $M^{2}_{B_{2}}$, respectively.}\label{3-point coupling}
\end{figure}

For the strong decay constants $f_{1}$ and $f_{2}$ of the decay $P_{c}(4312)\rightarrow J/\psi p$, similar analysis can be done and Figs.\ref{3-point coupling}(a) and \ref{3-point coupling}(b) exhibit the results with $M^{2}_{B_{2}}=0.8\mbox{GeV}^{2}$ and $M^{2}_{B_{1}}=3.8\mbox{GeV}^{2}$, respectively, both at $s_{J/\psi}=3.6^{2}\mbox{GeV}^{2}$ and $u_{N}=1.7\mbox{GeV}^{2}$. From Fig.\ref{3-point coupling}, we get the values of $f_{1}$ and $f_{2}$: $f_{1}=-0.486^{+0.076}_{-0.095}$ and $f_{2}=-0.571^{+0.077}_{-0.1}$. We list our values of the strong decay constants in Table.\ref{coupling}.

\begin{table}[htb]
\caption{Values of the strong decay constants.}\label{coupling}
\begin{tabular}{|c|c|}
  \hline
  Strong decay constant      &  Value   \\
  \hline
  {$g$}   &        $-0.419^{+0.019}_{-0.028}$\\
  \hline
  {$f_{1}$}  &      $-0.486^{+0.076}_{-0.095}$ \\
  \hline
  {$f_{2}$} &    $-0.571^{+0.077}_{-0.1}$ \\
  \hline
\end{tabular}
\end{table}

With all of the above parameters, the decay widths of $P_{c}(4312)\rightarrow \eta_{c} p$ and $P_{c}(4312)\rightarrow J/\psi p$ can be obtained. Using the transition matrix elements defined in Eq.(\ref{matrix element}) and following the standard method, one has
\begin{equation}
\Gamma(P_{c}(4312)\rightarrow \eta_{c} p)=\frac{g^{2}[(m_{P_{c}}+m_{N})^{2}-m^{2}_{\eta_{c}}]}{16\pi m^{3}_{P_{c}}}\sqrt{(m^{2}_{P_{c}}+m^{2}_{\eta_{c}}-m^{2}_{N})^{2}-4m^{2}_{P_{c}}m^{2}_{\eta_{c}}},
\end{equation}
\begin{eqnarray}
\Gamma(P_{c}(4312)\rightarrow J/\psi p)=&&\frac{(m_{p_{c}}+m_{N})^{2}-m^{2}_{J/\psi}}{16\pi m^{3}_{P_{c}}m^{2}_{J/\psi}(m_{p_{c}}+m_{N})^{2}}
[f^{2}_{1}(m_{p_{c}}+m_{N})^{2}(2m^{2}_{J/\psi}+(m_{P_{c}}-m_{N})^{2})\nonumber\\&&-6f_{1}f_{2}m^{2}_{J/\psi}(m^{2}_{P_{c}}-m^{2}_{N})+ f^{2}_{2}m^{2}_{J/\psi}(m^{2}_{J/\psi}+2(m_{P_{c}}-m_{N})^{2})]\nonumber\\&&\sqrt{(m^{2}_{P_{c}}+m^{2}_{N}-m^{2}_{J/\psi})^{2}-4m^{2}_{P_{c}}m^{2}_{N}}.
\end{eqnarray}
Substituting the values of the parameters involved in the above formulas, we find
\begin{eqnarray}
&&\Gamma(P_{c}(4312)\rightarrow \eta_{c} p)=5.54^{+0.75}_{-0.5}\mbox{MeV},
\nonumber\\
&&\Gamma(P_{c}(4312)\rightarrow J/\psi p)=1.67^{+0.92}_{-0.56}\mbox{MeV},
\end{eqnarray}
from which one has
\begin{eqnarray}
&&R\equiv\frac{\Gamma(P_{c}(4312)\rightarrow \eta_{c} p)}{\Gamma(P_{c}(4312)\rightarrow J/\psi p)}=3.32,\nonumber\\
&&\Gamma(P_{c}(4312)\rightarrow \eta_{c} p)+\Gamma(P_{c}(4312)\rightarrow J/\psi p)=7.21^{+1.67}_{-1.06}\mbox{MeV}.
\end{eqnarray}

In Refs.\cite{M.B.Voloshin,S.Sakai}, it was predicted that $R$ is $3$ based on the heavy quark spin symmetry. Obviously, our result is agreement with theirs taking into account the uncertainties. The sum of our partial decay widths is large, but still smaller than the total width of $P_{c}(4312)$, $\Gamma_{P_{c}(4312)}=9.8\pm2.7^{+3.7}_{-4.5}\mbox{MeV}$ reported by LHCb Collaboration \cite{lhcb1}.

\section{Conclusion}\label{sec4}

In the present work, the partial decay widths of $P_{c}(4312)\rightarrow \eta_{c} p$ and $P_{c}(4312)\rightarrow J/\psi p$ are studied via the method of QCD sum rule. As a starting point of our investigation, we assume the $P_{c}(4312)$ as a $\bar{D}\Sigma_{c}$ molecular state with $J^{P}=\frac{1}{2}^{-}$, which is reflected in the molecule-type interpolating current (\ref{Pc interpotating current}).

 The pole residue $\lambda_{P_{c}}$ of $P_{c}(4312)$ is an important parameter, which can be used as input parameter in the analyses of the electromagnetic properties and strong decays of $P_{c}(4312)$. Therefore, we firstly calculate the pole residue $\lambda_{P_{c}}$ by using two-point correlation function and get $\lambda_{P_{c}}=1.91^{+0.12}_{-0.13}\times10^{-3}\mbox{GeV}^{6}$. Then the strong decay constants are given by using three-point correlation functions and their values are $g=-0.419^{+0.019}_{-0.028}$, $f_{1}=-0.486^{+0.076}_{-0.095}$, and $f_{2}=-0.571^{+0.077}_{-0.1}$. With the numerical values of the strong decay constants, the partial decay widths to $\eta_{c} p$ and $J/\psi p$ are estimated to be $\Gamma(P_{c}(4312)\rightarrow \eta_{c} p)=5.54^{+0.75}_{-0.5}\mbox{MeV}$ and $\Gamma(P_{c}(4312)\rightarrow J/\psi p)=1.67^{+0.92}_{-0.56}\mbox{MeV}$, which are compatible with the total width of $P_{c}(4312)$ measured by LHCb Collaboration: $\Gamma_{P_{c}(4312)}=9.8\pm2.7^{+3.7}_{-4.5}\mbox{MeV}$. We also give the ratio $R$ of the decay width of $P_{c}(4312)\rightarrow \eta_{c} p$ to that of $P_{c}(4312)\rightarrow J/\psi p$, $R=3.32$, which is agreement with the values of Refs.\cite{M.B.Voloshin,S.Sakai}. In summary, it is reasonable to assign $P_{c}(4312)$ to be a $\bar{D}\Sigma_{c}$ molecular state with $J^{P}=\frac{1}{2}^{-}$.

\acknowledgments  This work was supported by the National
Natural Science Foundation of China under Contracts No.11675263 and 11405269.

\begin{appendix}
\section{The quark propagators}\label{appendix1}
The full quark propagators are
\begin{eqnarray}
 S^{q}_{ij}(x)=&&\frac{i \not\!{x}}{2\pi^{2}x^4}\delta_{ij}-\frac{m_{q}}{4\pi^2x^2}\delta_{ij}-\frac{\langle\bar{q}q\rangle}{12}\delta_{ij}
 +i\frac{\langle\bar{q}q\rangle}{48}m_{q}\not\!{x}\delta_{ij}-\frac{x^2}{192}\langle g_{s}\bar{q}\sigma Gq\rangle \delta_{ij}\nonumber\\
 &&+i\frac{x^2\not\!{x}}{1152}m_{q}\langle g_{s}\bar{q}\sigma Gq\rangle \delta_{ij}-i\frac{g_{s}t^{a}_{ij}G^{a}_{\mu\nu}}{32\pi^2x^2}(\not\!{x}\sigma^{\mu\nu}+\sigma^{\mu\nu}\not\!{x})+\cdots
 \end{eqnarray}
 for light quark, and
 \begin{eqnarray}
 S^{Q}_{ij}(x)=i\int\frac{d^{4}k}{(2\pi)^4}e^{-ikx}&&[\frac{\not\!{k}+m_{Q}}{k^2-m^{2}_{Q}}\delta_{ij}
 -\frac{g_{s}t^{a}_{ij}G^{a}_{\mu\nu}}{4}\frac{\sigma^{\mu\nu}(\not\!{k}+m_{Q})+(\not\!{k}+m_{Q})\sigma^{\mu\nu}}
 {(k^2-m^{2}_{Q})^{2}}\nonumber\\
 &&+\frac{\langle g^{2}_{s}GG\rangle}{12}\delta_{ij}m_{Q}\frac{k^2+m_{Q}\not\!{k}}{(k^2-m^{2}_{Q})^{4}}+\cdots]
 \end{eqnarray}
 for heavy quark. In these expressions, $t^{a}=\frac{\lambda^{a}}{2}$ and $\lambda^{a}$ are the Gell-Mann matrices, $g_{s}$ is the strong interaction coupling constant, and $i, j$ are color indices.

\section{The spectral densities}\label{appendix}
In this appendix, the spectral densities are given.

First, up to dimension-9 and $\alpha_{s}$ order, the spectral density $\rho_{1}(s)$ is
\begin{eqnarray}
\rho_{1}(s)=&&\rho^{0}_{1}(s)+\rho^{\langle\bar{q}q\rangle}_{1}(s)+\rho^{\langle g^{2}_{s}GG\rangle}_{1}(s)+\rho^{\langle g_{s}\bar{q}\sigma Gq\rangle}_{1}(s)+\rho^{\langle\bar{q}q\rangle^{2}}_{1}(s)\nonumber\\&&+\rho^{\langle\bar{q}q\rangle\langle g^{2}_{s}GG\rangle}_{1}(s)+\rho^{\langle\bar{q}q\rangle\langle g_{s}\bar{q}\sigma Gq\rangle}_{1}(s)+\rho^{\langle\bar{q}q\rangle^{3}}_{1}(s),
\end{eqnarray}
with
\begin{equation}
\rho^{0}_{1}(s)=-\frac{1}{20480\pi^{8}}\int^{a_{max}}_{a_{min}}\frac{da}{a^{4}}\int^{1-a}_{b_{min}}\frac{db}{b^{4}}(1-a-b)^{3}((a+b)m^{2}_{c}-abs)^{5},
\end{equation}
\begin{equation}
\rho^{\langle\bar{q}q\rangle}_{1}(s)=\frac{\langle\bar{q}q\rangle}{256\pi^{6}}m_{c}\int^{a_{max}}_{a_{min}}\frac{da}{a^{3}}\int^{1-a}_{b_{min}}\frac{db}{b^{2}}(1-a-b)^{2}((a+b)m^{2}_{c}-abs)^{3},
\end{equation}
\begin{eqnarray}
\rho^{\langle g^{2}_{s}GG\rangle}_{1}(s)=&&-\frac{\langle g^{2}_{s}GG\rangle}{24576\pi^{8}}m^{2}_{c}\int^{a_{max}}_{a_{min}}\frac{da}{a^{4}}\int^{1-a}_{b_{min}}\frac{db}{b^{4}}(a^{3}+b^{3})(1-a-b)^{3}((a+b)m^{2}_{c}-abs)^{2}\nonumber\\
&&-\frac{\langle g^{2}_{s}GG\rangle}{16384\pi^{8}}\int^{a_{max}}_{a_{min}}\frac{da}{a^{3}}\int^{1-a}_{b_{min}}\frac{db}{b^{3}}(2a+b)(1-a-b)^{2}((a+b)m^{2}_{c}-abs)^{3},
\end{eqnarray}
\begin{eqnarray}
\rho^{\langle g_{s}\bar{q}\sigma Gq\rangle}_{1}(s)=&&\frac{3\langle g_{s}\bar{q}\sigma Gq\rangle}{512\pi^{6}}m_{c}\int^{a_{max}}_{a_{min}}\frac{da}{a}\int^{1-a}_{b_{min}}\frac{db}{b^{2}}(1-a-b)((a+b)m^{2}_{c}-abs)^{2}\nonumber\\
&&-\frac{3\langle g_{s}\bar{q}\sigma Gq\rangle}{512\pi^{6}}m_{c}\int^{a_{max}}_{a_{min}}\frac{da}{a}\int^{1-a}_{b_{min}}\frac{db}{b^{3}}(1-a-b)^{2}((a+b)m^{2}_{c}-abs)^{2},
\end{eqnarray}
\begin{equation}
\rho^{\langle\bar{q}q\rangle^{2}}_{1}(s)=\frac{\langle\bar{q}q\rangle^{2}}{64\pi^{4}}\int^{a_{max}}_{a_{min}}\frac{da}{a}\int^{1-a}_{b_{min}}\frac{db}{b}((a+b)m^{2}_{c}-abs)^{2},
\end{equation}
\begin{eqnarray}
\rho^{\langle\bar{q}q\rangle\langle g^{2}_{s}GG\rangle}_{1}(s)=&&\frac{\langle\bar{q}q\rangle\langle g^{2}_{s}GG\rangle}{3072\pi^{6}}m^{3}_{c}\int^{a_{max}}_{a_{min}}\frac{da}{a^{2}}\int^{1-a}_{b_{min}}\frac{db}{b^{3}}(a^{3}+b^{3})(1-a-b)^{2}\nonumber\\
&&+\frac{\langle\bar{q}q\rangle\langle g^{2}_{s}GG\rangle}{1024\pi^{6}}m_{c}\int^{a_{max}}_{a_{min}}da\int^{1-a}_{b_{min}}\frac{db}{b^{3}}(1-a-b)^{2}((a+b)m^{2}_{c}-abs)\nonumber\\
&&+\frac{\langle\bar{q}q\rangle\langle g^{2}_{s}GG\rangle}{1024\pi^{6}}m_{c}\int^{a_{max}}_{a_{min}}\frac{da}{a}\int^{1-a}_{b_{min}}\frac{db}{b}(1-a-b)((a+b)m^{2}_{c}-abs),
\end{eqnarray}
\begin{eqnarray}
\rho^{\langle\bar{q}q\rangle\langle g_{s}\bar{q}\sigma Gq\rangle}_{1}(s)=&&-\frac{\langle\bar{q}q\rangle\langle g_{s}\bar{q}\sigma Gq\rangle}{128\pi^{4}}\int^{a_{max}}_{a_{min}}\frac{da}{a}\int^{1-a}_{b_{min}}db((a+b)m^{2}_{c}-abs)\nonumber\\
&&+\frac{\langle\bar{q}q\rangle\langle g_{s}\bar{q}\sigma Gq\rangle}{64\pi^{4}}\int^{a_{max}}_{a_{min}}da(m^{2}_{c}-a(1-a)s),
\end{eqnarray}
\begin{equation}
\rho^{\langle\bar{q}q\rangle^{3}}_{1}(s)=-\frac{\langle\bar{q}q\rangle^{3}}{24\pi^{2}}m_{c}\int^{a_{max}}_{a_{min}}da a,
\end{equation}
where $a_{max}=\frac{1+\sqrt{1-\frac{4m^{2}_{c}}{s}}}{2}$, $a_{min}=\frac{1-\sqrt{1-\frac{4m^{2}_{c}}{s}}}{2}$ and $b_{min}=\frac{am^{2}_{c}}{as-m^{2}_{c}}$.

Up to dimension-8 and $\alpha_{s}$ order, the explicit expressions of the spectral densities $\rho^{(3)}(s,u)$, $\rho^{(3)}_{1}(s,u)$ and $\rho^{(3)}_{2}(s,u)$ are
\begin{eqnarray}
\rho^{(3)}(s,u)=&&\frac{u^2\sqrt{s(s-4m^{2}_{c})}}{2048\pi^{6}}+\langle\bar{q}q\rangle\frac{m_{c}u\sqrt{s(s-4m^{2}_{c})}}{192\pi^{4}}\nonumber\\
&&+\langle g^{2}_{s}GG\rangle \frac{s u^{2}(3 m^{2}_{c}-s)}{18432\pi^{6}M^{4}_{B_{1}}\sqrt{s(s-4 m^{2}_{c})}}+\langle g^{2}_{s}GG\rangle \frac{u^{2}(m^{2}_{c} + s)}{12288\pi^{6} M^{2}_{B_{1}}\sqrt{s(s-4m^{2}_{c})}}\nonumber\\&&+\langle g^{2}_{s}GG\rangle \frac{m^{2}_{c} u (34 s + 3 u)}{24576 \pi^{6} s \sqrt{s (s - 4 m^{2}_{c})}}+\langle g^{2}_{s}GG\rangle \frac{\sqrt{s (s - 4 m^{2})}(s + 2 u)}{4096 \pi^{6} s}\nonumber\\&&+\langle g_{s}\bar{q}Gq\rangle\frac{m_{c} u}{384 \pi^{4}\sqrt{s(s-4m^{2}_{c})}}+\langle\bar{q}q\rangle^{2} \frac{\sqrt{s(s-4 m^{2}_{c})}\delta(u)}{48 \pi^{2}}\nonumber\\&&+\langle g^{2}_{s}GG\rangle\langle\bar{q}q\rangle \frac{s u (3 m^{2}_{c}-s)}{6912\pi^{4}M^{4}_{B_{1}}mc\sqrt{s (s-4 m^{2}_{c})}}+\langle g^{2}_{s}GG\rangle\langle\bar{q}q\rangle \frac{u (s - 2 m^{2}_{c})}{2304\pi^{4}M^{2}_{B_{1}}m_{c}\sqrt{s(s-4 m^{2}_{c})}}\nonumber\\&&-\langle g^{2}_{s}GG\rangle\langle\bar{q}q\rangle \frac{m_{c} \sqrt{s (s - 4 m^{2}_{c})} \delta(u)}{4608 \pi^{4} s}-\langle\bar{q}q\rangle\langle g_{s}\bar{q}Gq\rangle \frac{\sqrt{s(s-4 m^{2}_{c})} \delta(u)}{192 \pi^{2} M^{2}_{B_{2}}}\nonumber\\&&+\langle\bar{q}q\rangle\langle g_{s}\bar{q}Gq\rangle \frac{m^{2}_{c} \delta(u)}{36\pi^{2}\sqrt{s(s-4 m^{2}_{c})}}+\langle\bar{q}q\rangle\langle g_{s}\bar{q}Gq\rangle\frac{\sqrt{s(s - 4 m^{2}_{c})}\delta(u)}{96 \pi^{2}s},
\end{eqnarray}
\begin{eqnarray}
\rho^{(3)}_{1}(s,u)=&&\frac{u^{2}(s+2 m^{2}_{c})\sqrt{s(s-4 m^{2}_{c})}}{6144\pi^{6}s^{2}}-\langle g^{2}_{s}GG\rangle\frac{u^{2}(s-3 m^{2}_{c})}{36864 \pi^{6}M^{4}_{B_{1}}\sqrt{s(s-4 m^{2}_{c})}}\nonumber\\&&+\langle g^{2}_{s}GG\rangle\frac{m^{2}_{c}u^{2}}{36864 \pi^{6}M^{2}_{B_{1}}s\sqrt{s(s-4 m^{2}_{c})}}+\langle g^{2}_{s}GG\rangle\frac{19 m^{2}_{c}u}{73728\pi^{6}s\sqrt{s(s-4 m^{2}_{c})}}\nonumber\\&&+\langle g^{2}_{s}GG\rangle\frac{(s+2 m^{2}_{c})\sqrt{s(s-4 m^{2}_{c})}}{12288 \pi^{6}s^{2}}+\langle\bar{q}q\rangle\langle g_{s}\bar{q}Gq\rangle \frac{(s+2 m^{2}_{c})\sqrt{s (s-4 m^{2}_{c})}\delta(u)}{864 \pi^{2} M^{2}_{B_{2}} s^{2}}\nonumber\\&&+\langle\bar{q}q\rangle\langle g_{s}\bar{q}Gq\rangle \frac{m^{2}_{c}\delta(u)}{
 144 \pi^{2} s\sqrt{s (s - 4 m^{2}_{c})}},
\end{eqnarray}
and
\begin{eqnarray}
\rho^{(3)}_{2}(s,u)=&&-\frac{u^{2}\sqrt{s (s-4 m^{2}_{c})}(2 m^{2}_{c}+s)}{6144 \pi^{6} s}-\langle\bar{q}q\rangle \frac{m_{c} u\sqrt{s (s-4 m^{2}_{c})}}{192 \pi^{4} s}\nonumber\\&&+\langle g^{2}_{s}GG\rangle\frac{s u^{2} (s-3 mc^{2}_{c})}{36864 \pi^{6} M^{4}_{B_{1}}\sqrt{s (s-4 m^{2}_{c})}}+\langle g^{2}_{s}GG\rangle \frac{u^{2} (5 m^{2}_{c} - 2 s)}{36864 \pi^{6} M^{2}_{B_{1}}\sqrt{s (s - 4 m^{2})}}\nonumber\\&&+\langle g^{2}_{s}GG\rangle\frac{m^{2}_{c} u (u - 38 s)}{147456 \pi^{6} s\sqrt{s (s - 4 m^{2}_{c})}}-\langle g^{2}_{s}GG\rangle\frac{\sqrt{s (s - 4 m^{2}_{c})} (2 m^{2}_{c} + s)}{12288 \pi^{6} s}\nonumber\\&&+\langle g_{s}\bar{q}Gq\rangle \frac{m_{c} u (m^{2}_{c}-s)}{576 \pi^{4} s \sqrt{s (s-4 m^{2})}}+\langle g^{2}_{s}GG\rangle\langle\bar{q}q\rangle\frac{s u (s-3 m^{2}_{c})}{6912 \pi^{4} M^{4}_{B_{1}} m_{c}\sqrt{s (s-4 m^{2}_{c})}}\nonumber\\&&+\langle g^{2}_{s}GG\rangle\langle\bar{q}q\rangle\frac{u (10 m^{2}_{c} - 3 s)}{  6912 \pi^{4} M^{2}_{B_{1}} m_{c} \sqrt{s (s - 4 m^{2}_{c})}} + \langle g^{2}_{s}GG\rangle\langle\bar{q}q\rangle\frac{m_{c}\sqrt{s (s - 4 m^{2}_{c})}\delta(u)}{4608 \pi^{4}s}\nonumber\\&&-\langle\bar{q}q\rangle\langle g_{s}\bar{q}Gq\rangle \frac{\sqrt{s (s-4 m^{2}_{c})} (2 m^{2}_{c}+s) \delta(u)}{864 \pi^{2} M^{2}_{B_{2}} s}-\langle\bar{q}q\rangle\langle g_{s}\bar{q}Gq\rangle \frac{ m^{2}_{c} \delta(u)}{144 \pi^{2}\sqrt{s (s-4 m^{2}_{c})}}.
\end{eqnarray}
where $\delta(u)$ is the Dirac $\delta$-function.
\end{appendix}


\end{document}